\documentclass[journal]{IEEEtran}

\usepackage{color}
\usepackage{lineno}
\usepackage{cite}
\usepackage{verbatim}
\usepackage{pifont}
\usepackage{longtable}
\usepackage{tabularx}  
\usepackage{ltxtable}  
\usepackage{stfloats}
\usepackage{amsmath,amssymb,amsfonts}
\usepackage{algorithmic}
\usepackage{graphicx}
\usepackage{textcomp}
\usepackage{xcolor}
\usepackage{subfigure}
\usepackage{setspace}
\usepackage{svg}
\usepackage[ruled]{algorithm2e}
\usepackage{booktabs}
\usepackage{url}
\usepackage{diagbox}
\usepackage{xcolor}
\usepackage{multicol}
\usepackage{multirow}
\usepackage{tikz}
\usepackage[colorlinks=true,linkcolor=blue,citecolor=blue,urlcolor=blue]{hyperref}
\definecolor{lime}{HTML}{A6CE39}
\DeclareRobustCommand{\orcidicon}{%
    \begin{tikzpicture}
    \draw[lime, fill=lime] (0,0) 
    circle [radius=0.16] 
    node[white] {{\fontfamily{qag}\selectfont \tiny ID}};    \draw[white, fill=white] (-0.0625,0.095) 
    circle [radius=0.007];    \end{tikzpicture}
    \hspace{-2mm}}
\foreach \x in {A, ..., Z}{%
    \expandafter\xdef\csname orcid\x\endcsname{\noexpand\href{https://orcid.org/\csname orcidauthor\x\endcsname}{\noexpand\orcidicon}}
    }

\SetCommentSty{mycommfont}

\SetKwInput{KwInput}{Input}                
\SetKwInput{KwOutput}{Output}              

\begin{document}



\title{RadHARSimulator V1: Model-Based FMCW Radar Human Activity Recognition Simulator\\
\thanks{Manuscript received XXXXXXX XX, 2025; revised XXXXXXX XX, 2025; accepted XXXXXXX XX, 2025. Date of publication XXXXXXX XX, 2025; date of current version XXXXXXX XX, 2025.\par
My Bio: My name is Weicheng Gao. I'm a Ph.D. student from Beijing Institute of Technology. I’m majored and interested in mathematical and modeling theory research of signal processing, radar signal processing techniques, and AI for radar, apprenticed under professor Xiaopeng Yang. I’m currently dedicated in the field of Through-the-Wall Radar Human Activity Recognition. Looking forward to learning and collaborating with more like-minded teachers and mates. (e-mail: JoeyBG@126.com).\par
Digital Object Identifier 10.48550/arXiv.2509.XXXXX.\par}}

\author{Weicheng~Gao\orcidA{},~\IEEEmembership{Graduate~Student~Member,~IEEE}   
        \vspace{-0.4cm}
        }
        
\markboth{arXiv Preprint, September, 2025}%
{Shell \MakeLowercase{\textit{et al.}}: Bare Demo of IEEEtran.cls for IEEE Journals}

\maketitle

\begin{abstract}
Radar-based human activity recognition (HAR) is a pivotal research area for applications requiring non-invasive monitoring. However, the acquisition of diverse and high-fidelity radar datasets for robust algorithm development remains a significant challenge. To overcome this bottleneck, a model-based frequency-modulated continuous wave (FMCW) radar HAR simulator is developed. The simulator integrates an anthropometrically scaled $13$-scatterer kinematic model to simulate $12$ distinct activities. The FMCW radar echo model is employed, which incorporates dynamic radar cross-section (RCS), free-space or through-the-wall propagation, and a calibrated noise floor to ensure signal fidelity. The simulated raw data is then processed through a complete pipeline, including moving target indication (MTI), bulk Doppler compensation, and Savitzky-Golay denoising, culminating in the generation of high-resolution range-time map (RTM) and Doppler-time maps (DTMs) via both short-time Fourier transform (STFT) and Fourier synchrosqueezed transform (FSST). Finally, a novel neural network method is proposed to validate the effectiveness of the radar HAR. Numerical experiments demonstrate that the simulator successfully generates high-fidelity and distinct micro-Doppler signature, which provides a valuable tool for radar HAR algorithm design and validation. The installer of this simulator is released at: \href{https://github.com/JoeyBGOfficial/RadHARSimulatorV1-Model-Based-FMCW-Radar-Human-Activity-Recognition-Simulator}{Github/JoeyBGOfficial/RadHARSimulatorV1}.\par
\end{abstract}


\IEEEpeerreviewmaketitle
\section{Introduction}
\IEEEPARstart{I}{n} recent years, significant attention is directed toward radar-based human activity recognition (HAR) due to its non-intrusive nature and robustness in diverse environmental conditions \cite{Main1, Main2}. Frequency-modulated continuous wave (FMCW) radar, in particular, is recognized for its ability to capture detailed micro-Doppler signature, enabling the identification of complex human motions \cite{Main3, Main4}. Research has been conducted on free-space and through-the-wall detection, applicable to various activity categories and human body, among other scenarios \cite{Gao1, Gao2, Gao3, Gao4, Gao5, Gao6}. The goal is to develop a systematic approach for radar HAR systems with high practical value.\par
\begin{figure}
    \centering
    \includegraphics[width=0.47\textwidth]{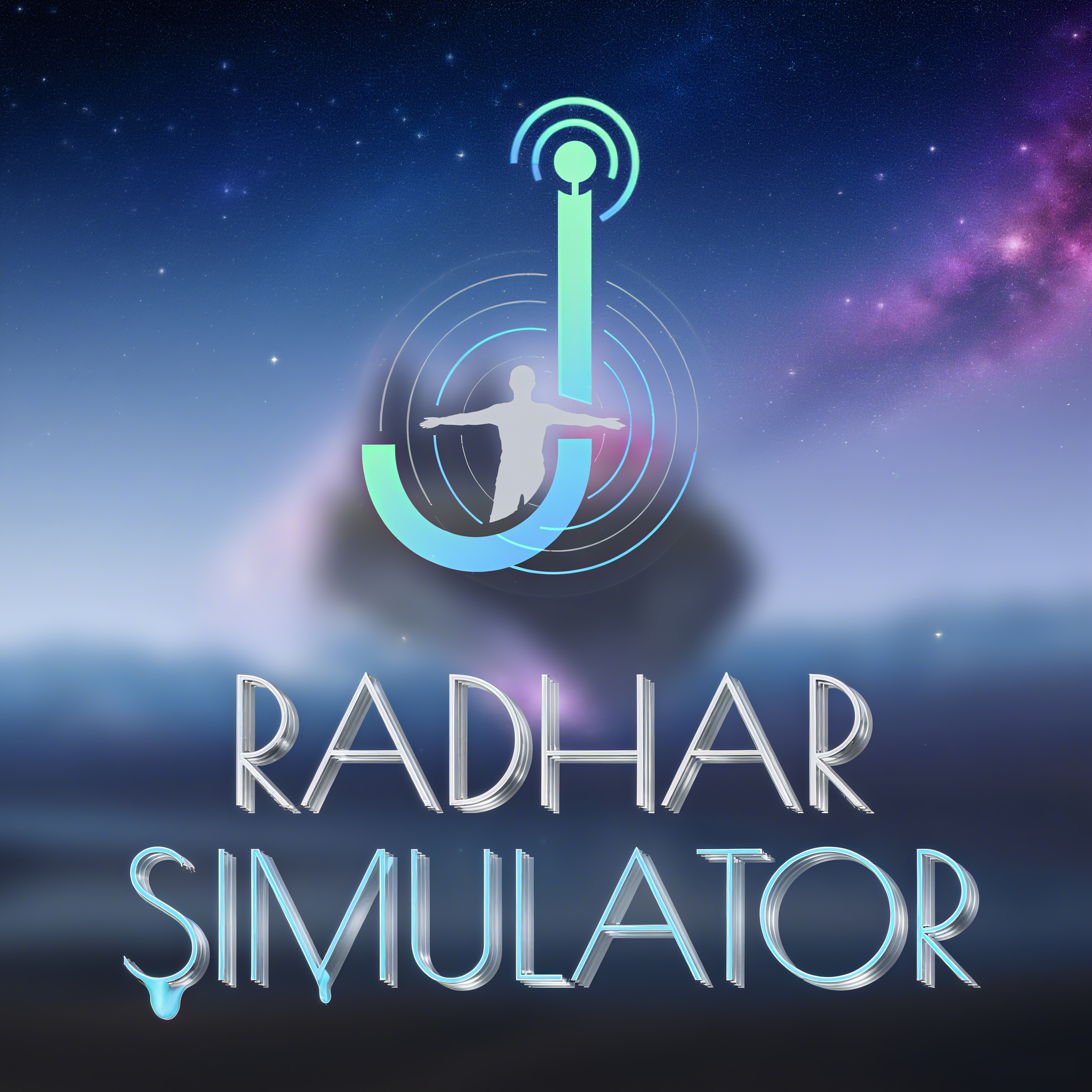}
    \caption{Splash screen of RadHARSimulator V1.}
    \label{Splash_Screen}
    \vspace{-0.4cm}
\end{figure}\par
\begin{figure*}[!ht]
    \centering
    \fcolorbox{lightgray}{white}{\includegraphics[width=0.98\textwidth]{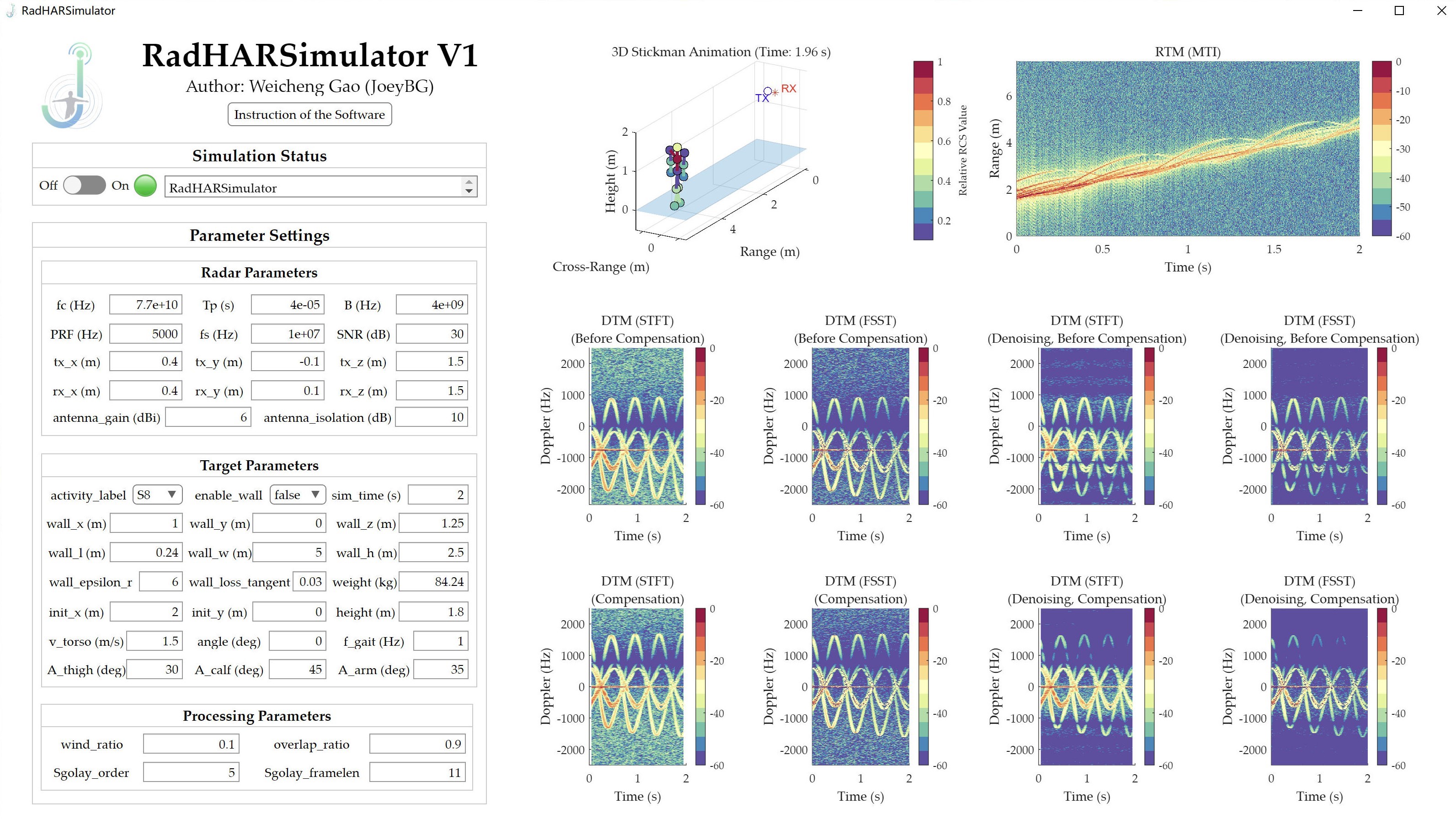}}
    \caption{The software interface of RadHARSimulator V1.\label{Software_Interface}}
    \vspace{-0.2cm}
\end{figure*}\par
Remarkable research achievements have been made in the field of radar HAR. Deep learning techniques for radar-based HAR were systematically surveyed \cite{li2019survey}. To improve feature extraction and recognition accuracy with limited data, a novel range–time–Doppler map and a range-distributed convolutional neural network (CNN) were introduced\cite{kim2020rtdcnn}. A real-time radar HAR system deployable on edge devices was developed \cite{gianoglio2024edge}. For robust and accurate recognition, a noninvasive HAR system using FMCW radar was proposed \cite{yu2022mmwave}. A semisupervised transfer learning algorithm combining domain adaptation and semantic transfer was presented, enabling effective radar HAR with sparsely labeled data \cite{li2021semisupervised}. For efficient HAR on mobile-edge devices, Mobile-RadarNet, a lightweight CNN architecture using depthwise and pointwise convolutions, was designed \cite{zhu2021mobile}. A lightweight hybrid vision transformer (ViT) network combining convolution and self-attention was proposed to achieve high accuracy and efficiency in embedded scenarios \cite{huan2023vit}. To improve recognition accuracy, a mixed CNN architecture that fused features from multiple time-frequency spectrograms was introduced \cite{tang2021mcnn}, and a one-dimensional (1D) dense attention neural network with time and frequency attention branches was designed to enhance feature utilization \cite{lai2021dan}. Concurrently, an efficient attention-based CNN that decoupled Doppler and temporal features was proposed to improve accuracy and real-time performance \cite{huan2023attention}. A CEEMD-ES multistatic radar selection method was introduced to optimize radar choice and feature extraction \cite{zhang2021ceemd}. To reduce reliance on large experimental datasets, a fully simulation-based framework for HAR system design was proposed \cite{waqar2024simulation}, while a distributed radar sensor system and a hybrid neural network were developed to address detection blindness in multi-radar scenarios \cite{zheng2024distributed}. A lightweight hardware accelerator was also designed to significantly improve energy-resource efficiency \cite{long2024fpga}. Furthermore, a two-stage domain adaptation method was proposed to enable HAR with limited measurement data by using simulated spectrograms \cite{yang2023domain}. A two-stream 1D-CNN with cross-channel operation were combined for micro-Doppler data to achieve superior accuracy \cite{tan2024microdoppler}, and a spectro-temporal network was introduced for joint temporal and frequency modeling \cite{luo2023stnet}. A source hypothesis transfer learning framework was developed for cross-environment HAR without source data \cite{cao2021crossenv}, and finally, the applications and challenges of the field were summarized to provide comprehensive guidance for future research \cite{luo2024challenges}. However, the difficulty in data acquisition remains a significant challenge in this domain.\par
The motivation for developing a simulation tool stems from the limitations of experimental radar data collection. Real-world data acquisition is often constrained by high costs, limited access to diverse testers, and challenges in replicating specific scenarios, such as through-the-wall detection or complex activity transitions. The simulator is proposed to overcome these barriers by providing a controlled environment where radar parameters, human kinematics, and environmental factors can be systematically varied. By generating synthetic radar echoes that incorporate realistic physical models, such as bistatic radar configurations and wall attenuation effects, a simulation platform is enabled to support the design and validation of HAR algorithms without the need for extensive physical experiments.\par
The RadHARSimulator V1, shown in Fig. \ref{Software_Interface}, is equipped with a comprehensive set of features to model and analyze human activities. A Boulic-Thalmann inspired kinematic model representing the human body as $13$ scatterers with anthropometrically scaled dimensions is employed \cite{Boulic-Thalmann}, allowing for realistic simulation of $12$ distinct activities \cite{SimHumalator}, ranging from stationary postures to complex transitions like walking to sitting and walking to falling. Bistatic FMCW radar echoes, including free-space or through-the-wall effects and additive white Gaussian noise (AWGN), are generated to mimic real-world conditions. The simulator produces a three-dimensional (3D) stickman animation, where joint colors reflect radar cross-section (RCS) values, alongside visualizations of the range-time map (RTM) and Doppler-time maps (DTMs) using short-time Fourier transform (STFT) and Fourier synchrosqueezed transform (FSST). Signal processing techniques, including moving target indication (MTI), bulk Doppler compensation, and Savitzky-Golay denoising, are integrated to enhance the quality of the generated data. Entropy metrics for RTM and DTMs are computed to provide quantitative insights into signal characteristics. Furthermore, based on a modified star-shaped operation, a novel neural network architecture is proposed to validate the effectiveness of the image data generated by the simulator and the feasibility of radar HAR. Numerical experiments demonstrate that the simulator successfully generates high-fidelity and distinct micro-Doppler signature, which provides a valuable tool for radar HAR algorithm design and validation. The simulator is available for free download on the GitHub platform.\par
The rest of this paper is organized as follows. Section II presents the modeling theory, detailing the kinematic and radar echo model. Section III describes the signal processing method, including the generation of RTM and DTMs. Section IV discusses the design of the proposed neural network. Section V gives numerical experiments to evaluate the simulator’s performance. Finally, Section VI concludes the paper with a summary of contributions and directions for future work.\par

\section{Modeling Theory}
In this section, a hierarchical multi-point kinematic model is constructed to represent human motion across $12$ distinct activities. Subsequently, the FMCW radar echo model is derived from foundational principles including both free-space and through-the-wall detection scenarios.\par

\subsection{Human Kinematic Model}
\begin{figure}
    \centering
    \includegraphics[width=0.48\textwidth]{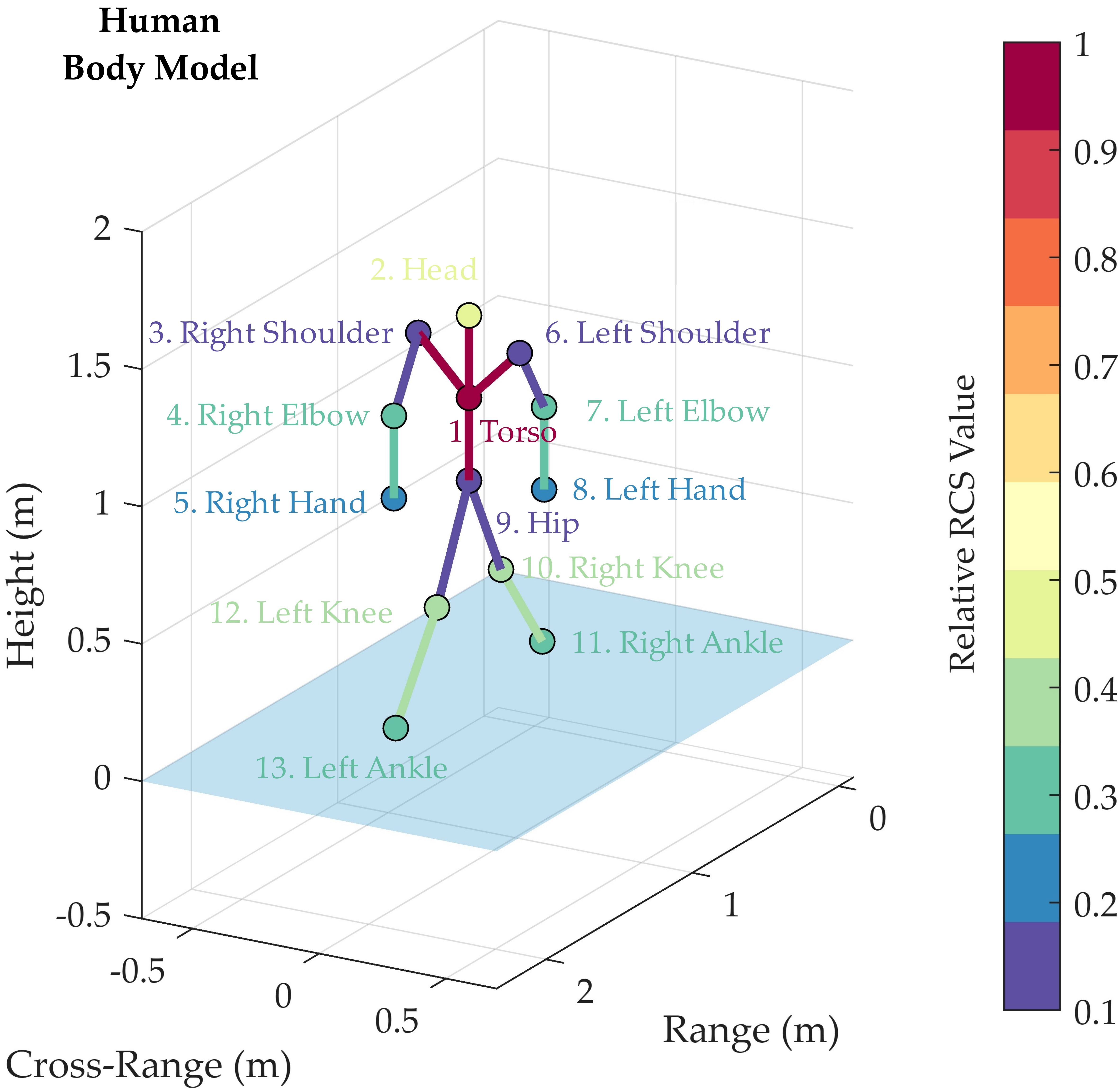}
    \caption{The $13$-joint human kinematic modeling.\label{Human_Modeling}}
    \vspace{-0.4cm}
\end{figure}

As shown in Fig. \ref{Human_Modeling}, the human form is modeled as a kinematic chain composed of $N=13$ discrete point scatterers, which correspond to major anatomical joints \cite{Boulic-Thalmann}. A hierarchical parent-child structure is employed, in which the global position of the entire system is dictated by the trajectory of the "Torso" as the root node.\par
Let the Cartesian coordinate vectors for the radar transmitter and receiver be denoted by $\mathbf{p}_{\mathrm{TX}} \in \mathbb{R}^3$ and $\mathbf{p}_{\mathrm{RX}} \in \mathbb{R}^3$, respectively. Define the x, y, z axes to represent range, cross-range, and height direction, respectively. The position of the $i^\mathrm{th}$ scatterer at a given slow-time $t$ is represented by the vector $\mathbf{p}_i(t) \in \mathbb{R}^3$. The position of the torso ($i=1$ node) is defined by its global trajectory. For any child scatterer $i$ with a parent $p(i)$, its position is computed by adding a rotated link vector to the parent's position:

\vspace{-0.2cm}
\begin{equation}
\mathbf{p}_i(t) = \mathbf{p}_{p(i)}(t) + \mathbf{R}_{\mathrm{tot}, i}(t) \mathbf{l}_i^0,
\end{equation}
where $\mathbf{l}_i^0$ is the static link vector in the parent's local frame, and $\mathbf{R}_{\mathrm{tot}, i}(t)$ is the total rotation matrix. This rotation is a composition of a global rotation matrix, $\mathbf{R}_{\mathrm{global}}(t)$, and a chain of local rotation matrices, $\mathbf{R}_{\mathrm{local}, i}(t)$, corresponding to each joint $i$ in the kinematic chain from the root to the current link. Local joint motion is modeled primarily as a rotation around the local y-axis by a time-varying angle $\theta_i(t)$:

\vspace{-0.2cm}
\begin{equation}
\mathbf{R}_{\mathrm{local}, i}(\theta_i(t)) = 
\begin{pmatrix}
\cos \theta_i(t) & 0 & \sin \theta_i(t) \\
0 & 1 & 0 \\
-\sin \theta_i(t) & 0 & \cos \theta_i(t)
\end{pmatrix}.
\end{equation}\par
Based on the SimHumalator developed at University College London \cite{SimHumalator}, $12$ common indoor human activities are defined: $S1$, Stationary; $S2$, Punching; $S3$, Kicking; $S4$, Grabbing; $S5$, Sitting Down; $S6$, Standing Up; $S7$, Body Rotating; $S8$, Walking; $S9$, Sitting to Walking; $S10$, Walking to Sitting; $S11$, Falling to Walking; $S12$, Walking to Falling. Below, the motion models for each activity are derived.\par

\textbf{$\mathbf{S1}$ - Stationary:}\par
In this activity, the human remains completely still in an upright standing posture. All body segments are static relative to one another and to the global coordinate system. The velocity of torso $\mathbf{v}_\mathrm{torso}$ is:

\vspace{-0.2cm}
\begin{equation}
\mathbf{v}_\mathrm{torso}=\mathbf{0}^\top.
\end{equation}\par
Therefore, the torso's position remains constant at its initial position for time $t$:

\vspace{-0.2cm}
\begin{equation}
\mathbf{p}_1(t) \equiv \mathbf{p}_1(0) = [x_0, y_0, h_{\mathrm{torso}}]^{\top},
\end{equation}
where $x_0,y_0$ are initial range and cross-range coordinates of torso, and $h_{\mathrm{torso}}$ is the height of torso to the ground.\par
As there is no limb movement, all local joint rotation angles are zero for all time:

\vspace{-0.2cm}
\begin{equation}
\theta_i(t) = 0, \quad \forall i \in \{1, \ldots, 13\}.
\end{equation}\par

\textbf{$\mathbf{S2}$ - Punching:}\par
In this activity, the human stands in place and performs alternating punches with the right and left arms. The torso and legs remain stationary. The motion is confined to the arms, originating at the shoulders. The punches are periodic. The torso's position for time $t$ is:

\vspace{-0.2cm}
\begin{equation}
\mathbf{p}_1(t) = [x_0, y_0, h_{\mathrm{torso}}]^{\top}.
\end{equation}\par
Define the arm swing frequency as equal to the uniform gait frequency parameter $f_g$. Based on the sinusoidal oscillation model, the rotation angle of the right elbow is:

\vspace{-0.2cm}
\begin{equation}
\theta_4(t) = -\frac{\pi}{4} - A_{\mathrm{arm}} \sin(2\pi f_g t),
\end{equation}
where $A_{\mathrm{arm}}$ is the maximum range of motion of the elbow.\par
The right hand and right elbow rotate with a difference of $\pi/4$:

\vspace{-0.2cm}
\begin{equation}
\theta_5(t) = -\frac{\pi}{2} - A_{\mathrm{calf}} \cos(2\pi f_g t),
\end{equation}
where $A_{\mathrm{calf}}$ is used to measure the maximum range of motion of both hands.\par
The left elbow and left hand are mirrored with respect to the right elbow and right hand nodes:

\vspace{-0.2cm}
\begin{equation}
\theta_7(t) = -\frac{\pi}{4} + A_{\mathrm{arm}} \sin(2\pi f_g t),
\end{equation}

\vspace{-0.2cm}
\begin{equation}
\theta_8(t) = -\frac{\pi}{2} + A_{\mathrm{calf}} \cos(2\pi f_g t),
\end{equation}\par
All other joints remain a non-rotating state:

\vspace{-0.2cm}
\begin{equation}
\theta_i(t) = 0, \quad \forall i \in \{1, 2, 3, 6, 9, \ldots, 13\}.
\end{equation}\par

\textbf{$\mathbf{S3}$ - Kicking:}\par
In this activity, the human stands in place and performs a repetitive kicking motion with the right leg. The torso, arms, and left leg remain stationary. The motion is confined to the right leg and is periodic with the frequency of $f_g$. The torso's position for time $t$ is:

\vspace{-0.2cm}
\begin{equation}
\mathbf{p}_1(t) = [x_0, y_0, h_{\mathrm{torso}}]^{\top}.
\end{equation}\par
The thigh swings forward and backward from the hip:

\vspace{-0.2cm}
\begin{equation}
\theta_{10}(t) = A_{\mathrm{thigh}} \sin(2\pi f_g t),
\end{equation}
where $A_{\mathrm{thigh}}$ is the maximum range of motion of the thigh.\par
The lower leg also swings at the knee with the phase difference of $\pi/6$:

\vspace{-0.2cm}
\begin{equation}
\theta_{11}(t) = \frac{\pi}{6} + A_{\mathrm{calf}} \sin(2\pi f_g t),
\end{equation}
where $A_{\mathrm{calf}}$ is used to measure the maximum range of motion of the calf.\par
All other joints remain a non-rotating state:

\vspace{-0.2cm}
\begin{equation}
\theta_i(t) = 0, \quad \forall i \in \{1, \ldots, 9, 12, 13\}.
\end{equation}\par

\textbf{$\mathbf{S4}$ - Grabbing:}\par
In this activity, the human bends at the waist to grab something from the floor and then returns to an upright position. The feet remain planted, and the motion is dominated by the rotation of the torso with the frequency of $f_g$. To maintain balance and posture, other joints must counter-rotate. The torso's position for time $t$ is:

\vspace{-0.2cm}
\begin{equation}
\mathbf{p}_1(t) = [x_0, y_0, h_{\mathrm{torso}}]^{\top}.
\end{equation}\par
To model a smooth transition from an angle of $0$ to $\pi$ radians and back of torso, a raised cosine function is ideal: 

\vspace{-0.2cm}
\begin{equation}
\theta_1(t) = \frac{\pi}{2}(1 - \cos(2\pi f_g t)).
\end{equation}\par
To keep the legs vertical as the torso bends forward, the hip joint must rotate backward by the same amount as the torso rotates forward: 

\vspace{-0.2cm}
\begin{equation}
\theta_9(t) = \theta_4(t) = -\theta_1(t) = -\frac{\pi}{2}(1 - \cos(2\pi f_g t)).
\end{equation}\par
As the torso rotates, all child joints inherit this rotation. Specific counter-rotations at the shoulders and elbows to simulate the arms reaching down are defined:

\vspace{-0.2cm}
\begin{equation}
\theta_7(t) = -\frac{\pi}{3} (1 - \cos(2\pi f_g t)),
\end{equation}

\vspace{-0.2cm}
\begin{equation}
\theta_8(t) = -\frac{\pi}{12} (1 - \cos(2\pi f_g t)).
\end{equation}\par
All other joints remain a non-rotating state:

\vspace{-0.2cm}
\begin{equation}
\theta_i(t) = 0, \quad \forall i \in \{2, 3, 5, 6, 10, \ldots, 13\}.
\end{equation}\par

\textbf{$\mathbf{S5}$ - Sitting Down:}\par
In this activity, the human transitions from a standing position to a seated position. The motion is a controlled squat. The feet remain stationary on the ground. The entire maneuver unfolds over a duration $T_{\mathrm{sit}}$. A progress variable $P(t)$ that goes smoothly from $0$  to $1$ is defined:

\vspace{-0.2cm}
\begin{equation}
P(t) = 0.5 \left(1 - \cos\left(\pi \cdot \min\left(1, \frac{t}{T_{\mathrm{sit}}}\right)\right)\right).
\end{equation}\par
As the person sits, the knees and hips bend from $0$ to $\pi/2$. The rotation angles are directly proportional to the progress $P(t)$. The hip rotates backward relative to the torso, and the knee rotates forward relative to the thigh to fold the leg:

\vspace{-0.2cm}
\begin{equation}
\theta_9(t) = -\frac{\pi}{2} P(t),
\end{equation}

\vspace{-0.2cm}
\begin{equation}
\theta_{10}(t) = \theta_{12}(t) = \frac{\pi}{2} P(t).
\end{equation}\par
Consider the right-angle triangle formed by the hip, knee, and ankle when seated. The torso must shift backward horizontally and drop vertically. The horizontal shift of the hip relative to the ankle is $L_{\mathrm{thigh}}\sin(\theta_{10}(t))$, and the vertical drop is $L_{\mathrm{thigh}}(1 - \cos(\theta_{10}(t)))$, where $L_{\mathrm{thigh}}$ is the length of thigh. Therefore, the torso's position for time $t$ is:

\vspace{-0.2cm}
\begin{equation}
\mathbf{p}_1(t) = \begin{pmatrix} x_0 - L_{\mathrm{thigh}}\sin(\frac{\pi}{2}P(t)) \\ y_0 \\ h_{\mathrm{torso}} - L_{\mathrm{thigh}}(1-\cos(\frac{\pi}{2}P(t))) \end{pmatrix}.
\end{equation}\par
All other joints remain a non-rotating state:

\vspace{-0.2cm}
\begin{equation}
\theta_i(t) = 0, \quad \forall i \in \{1, \ldots, 8, 11, 13\}.
\end{equation}\par

\textbf{$\mathbf{S6}$ - Standing Up:}\par
In this activity, the human transitions from a seated to a standing position. The logic is identical to $S5$, but the progress is reversed. The progress function $P(t)$ from $S5$ is used to define $P_{\mathrm{stand}}(t) = 1 - P(t)$. The joint angles and torso displacements now evolve from their seated state back to zero. The angles of legs go from $\pm \pi/2$ back to $0$:

\vspace{-0.2cm}
\begin{equation}
\theta_9(t) = -\frac{\pi}{2}P_{\mathrm{stand}}(t) = -\frac{\pi}{2} (1 - P(t)),
\end{equation}

\vspace{-0.2cm}
\begin{equation}
\theta_{10}(t) = \theta_{12}(t) = \frac{\pi}{2}P_{\mathrm{stand}}(t) = \frac{\pi}{2} (1 - P(t)).
\end{equation}\par
The torso moves from its lowered and shifted position back to the initial standing position. Therefore, the torso's position for time $t$ is:

\vspace{-0.2cm}
\begin{equation}
\mathbf{p}_1(t) = \begin{pmatrix} x_0 - L_{\mathrm{thigh}}\sin(\frac{\pi}{2}(1-P(t))) \\ y_0 \\ h_{\mathrm{torso}} - L_{\mathrm{thigh}}(1-\cos(\frac{\pi}{2}(1-P(t)))) \end{pmatrix}.
\end{equation}\par
All other joints remain a non-rotating state:

\vspace{-0.2cm}
\begin{equation}
\theta_i(t) = 0, \quad \forall i \in \{1, \ldots, 8, 11, 13\}.
\end{equation}\par
\begin{table}
\begin{center}
\caption{Structure and Properties of Human Body Model$^{*}$.\label{Human Body Model}}
\vspace{-0.0cm}
\resizebox{0.48\textwidth}{!}{
\begin{tabular}{cccc}
\hline\hline
\textbf{Joint} & \textbf{Parent} & \textbf{Relative Position [x,y,z]} ($m$) & \textbf{RCS} ($m^2$) \\
\hline
Torso & / & $[0, 0, 0]$ & $1.0$ \\
Head & Torso & $[0, 0, 0.3]$ & $0.5$ \\
Right Shoulder & Torso & $[0, -0.2, 0.2]$ & $0.1$ \\
Right Elbow & Right Shoulder & $[0, 0, -0.3]$ & $0.3$ \\
Right Hand & Right Elbow & $[0, 0, -0.3]$ & $0.2$ \\
Left Shoulder & Torso & $[0, 0.2, 0.2]$ & $0.1$ \\
Left Elbow & Left Shoulder & $[0, 0, -0.3]$ & $0.3$ \\
Left Hand & Left Elbow & $[0, 0, -0.3]$ & $0.2$ \\
Hip & Torso & $[0, 0, -0.3]$ & $0.1$ \\
Right Knee & Hip & $[0, 0, -0.45]$ & $0.4$ \\
Right Ankle & Right Knee & $[0, 0, -0.45]$ & $0.3$ \\
Left Knee & Hip & $[0, 0, -0.45]$ & $0.4$ \\
Left Ankle & Left Knee & $[0, 0, -0.45]$ & $0.3$ \\
\hline\hline
\end{tabular}
}
\end{center}
\footnotesize $^{*}$ All values of relative position and RCS are defined under $1.8~m$ height and $84.24\mathrm{~kg}$ weight human. Distances between joints can be scaled proportionally based on height, while the RCS of each joint can be scaled proportionally by taking the square root of the product of height and weight.\\
\vspace{-0.4cm}
\end{table}\par

\textbf{$\mathbf{S7}$ - Body Rotating:}\par
In this activity, the human stands in place and twists their upper body left and right. The body rotates as a rigid unit around the vertical z-axis passing through the center of the torso with the frequency of $f_g$. The torso's position is static:

\vspace{-0.2cm}
\begin{equation}
\mathbf{p}_1(t) = [x_0, y_0, h_{\mathrm{torso}}]^{\top}.
\end{equation}\par
The motion can be easily modeled in the global rotation matrix $\mathbf{R}_{\mathrm{global}}(t)$. A time-varying rotation angle $\phi(t)$ around the z-axis is defined. A sine wave is used to model the oscillatory twisting motion:

\vspace{-0.2cm}
\begin{equation}
\phi(t) = A_{\mathrm{turn}} \sin(2\pi f_g t),
\end{equation}
where $A_{\mathrm{turn}}$ is the maximum body rotating range. This angle is used to formulate the global rotation for time $t$:

\vspace{-0.2cm}
\begin{equation}
\mathbf{R}_{\mathrm{global}}(t) = 
\begin{pmatrix}
\cos \phi(t) & -\sin \phi(t) & 0 \\
\sin \phi(t) & \cos \phi(t) & 0 \\
0 & 0 & 1
\end{pmatrix}.
\end{equation}\par
This matrix is then applied to all link vectors during the kinematic calculation, rotating the entire body structure.\par
The rotation angle of each joint can be set to $0$:

\vspace{-0.2cm}
\begin{equation}
\theta_i(t) = 0, \quad \forall i.
\end{equation}\par

\textbf{$\mathbf{S8}$ - Walking:}\par
This is the standard, steady-state walking gait activity. The torso moves forward with a constant velocity. Arm and leg movements are periodic with the frequency of $f_g$ and coordinated in antiphase. The torso moves with constant velocity $\mathbf{v}_{\mathrm{torso}} = [v_x, v_y, 0]^{\top}$. The torso's position for time $t$ is:

\vspace{-0.2cm}
\begin{equation}
\mathbf{p}_1(t) = [x_0 + v_x t, y_0 + v_y t, h_{\mathrm{torso}}]^{\top}.
\end{equation}\par
The right and left legs swing in antiphase, which is modeled with sine waves of opposite signs:

\vspace{-0.2cm}
\begin{equation}
\theta_{10}(t) = A_{\mathrm{thigh}} \sin(2\pi f_g t),
\end{equation}

\vspace{-0.2cm}
\begin{equation}
\theta_{12}(t) = -A_{\mathrm{thigh}} \sin(2\pi f_g t).
\end{equation}\par
The lower leg's motion is not perfectly in sync with the thigh. A phase lag of $\pi/4$ is introduced to model the natural bend and extension during the swing phase:

\vspace{-0.2cm}
\begin{equation}
\theta_{11}(t) = A_{\mathrm{calf}} \sin(2\pi f_g t + \pi/4),
\end{equation}

\vspace{-0.2cm}
\begin{equation}
\theta_{13}(t) = -A_{\mathrm{calf}} \sin(2\pi f_g t + \pi/4).
\end{equation}\par
Arms swing to counterbalance the legs. They are in antiphase with their corresponding legs with a phase shift of $\pi$ to achieve this:

\vspace{-0.2cm}
\begin{equation}
\theta_4(t) = A_{\mathrm{arm}} \sin(2\pi f_g t + \pi),
\end{equation}

\vspace{-0.2cm}
\begin{equation}
\theta_7(t) = -A_{\mathrm{arm}} \sin(2\pi f_g t + \pi).
\end{equation}\par
All other joints remain a non-rotating state:

\vspace{-0.2cm}
\begin{equation}
\theta_i(t) = 0, \quad \forall i \in \{1, 2, 3, 5, 6, 8, 9\}.
\end{equation}\par

\textbf{$\mathbf{S9}$ - Sitting to Walking:}\par
In this activity, the human starts from a seated position, stands up, and then begins to walk. The motion is segmented into two distinct phases: (1) Standing up, and (2) Walking. The kinematics for each phase are drawn from the models above and stitched together. Let the standing up begin at $0$ and end at $t_w$. Let walking begin at $t_w$.\par
Phase 1: Standing Up ($0 \le t < t_w$): The kinematics of $S6$ are applied. The torso moves from the seated position back to the upright, stationary position.\par
Phase 2: Walking ($t \ge t_w $): The kinematics of $S8$ are applied. The time variable is shifted: $t' = t - t_w$. The torso begins moving with constant velocity from the position it reached at the end of Phase 2.\par

\textbf{$\mathbf{S10}$ - Walking to Sitting:}\par
In this activity, the human is walking, then stops and sits down. A two-phase composite activity is modeled. Let the transition to sitting begin at time $t_{s}$.\par
Phase 1: Walking ($0 \le t < t_{s}$): The human follows the $S8$ model. The torso moves with constant velocity.\par
Phase 2: Sitting Down ($t \ge t_{s}$): The human follows the $S5$ model. The time variable is shifted: $t' = t - t_{\mathrm{s}}$. The initial position for the sitting motion is the torso's final position at the end of the walking phase. The leg kinematics smoothly transition from walking oscillations to the controlled bending of the sitting maneuver.\par

\textbf{$\mathbf{S11}$ - Falling to Walking:}\par
In this activity, the human begins lying on the floor, gets up, and starts walking. A two-phase activity is modeled. The getting up is modeled as a whole-body rotation from parallel to vertical. Let getting up start at $0$ and walking start at $t_w$.\par
Phase 1: Getting Up ($0 \le t < t_w$): The body model has a constant global rotation of $\pi/2$ around the local y-axis. The torso's height, $h_{\mathrm{torso}}$ is set to a small value around $0.15~m$. The transition is modeled by smoothly decreasing the global rotation angle $\theta_{\mathrm{global}}(t)$ from $\pi/2$ back to $0$, while simultaneously increasing the torso's height from its low value back to the normal $h_{\mathrm{torso}}$. This is controlled by the progress function $P(t)$:

\vspace{-0.2cm}
\begin{equation}
\theta_{\mathrm{global}}(t) = \frac{\pi}{2}(1 - P(t)),
\end{equation}

\vspace{-0.2cm}
\begin{equation}
z_1(t) = 0.15 + (h_{\mathrm{torso}} - 0.15) P(t),
\end{equation}
where $z_1(t)$ is the height of torso for time $t$.\par
Phase 3: Walking ($t \ge t_w$): The $S8$ model is engaged with a shifted time $t'=t-t_w$.\par
\begin{table*}
\begin{center}
\caption{Rotation Functions for Joints Across Different Activities$^{*}$.\label{Rotation Functions}}
\vspace{-0.5cm}
\resizebox{\textwidth}{!}{
\begin{tabular}{ccc}
\hline\hline
\textbf{Activity Label} & \textbf{Joint Name} & \textbf{Rotation Function $\theta(t)$} \\
\hline
$S1$: Stationary
& Torso, Head, Shoulders, Elbows, Hands, Hip, Knees, Ankles & $0$ \\
\hline
\multirow{5}{*}{$S2$: Punching} 
& Torso, Head, Right Shoulder, Left Shoulder, Hip, Knees, Ankles & $0$ \\
& Right Elbow & $ - \frac{\pi}{4} - A_{\text{arm}}\sin(2\pi f_g t)$ \\
& Right Hand & $ - \frac{\pi}{2} - A_{\text{calf}}\cos(2\pi f_g t)$ \\
& Left Elbow & $ - \frac{\pi}{4} + A_{\text{arm}}\sin(2\pi f_g t)$ \\
& Left Hand & $ - \frac{\pi}{2} + A_{\text{calf}}\cos(2\pi f_g t)$ \\
\hline
\multirow{3}{*}{$S3$: Kicking} 
& Torso, Head, Shoulders, Elbows, Hands, Hip, Left Knee, Left Ankle & $0$ \\
& Right Knee & $A_{\text{thigh}}\sin(2\pi f_g t)$ \\
& Right Ankle & $\frac{\pi}{6} + A_{\text{calf}}\sin(2\pi f_g t)$ \\
\hline
\multirow{6}{*}{$S4$: Grabbing} 
& Head, Shoulders, Right Hand, Knees, Ankles & $0$ \\
& Torso & $\frac{\pi}{2}(1 - \cos(2\pi f_g t))$ \\
& Right Elbow & $-\frac{\pi}{2}(1 - \cos(2\pi f_g t))$ \\
& Left Elbow & $-\frac{\pi}{3}(1 - \cos(2\pi f_g t))$ \\
& Left Hand & $-\frac{\pi}{12}(1 - \cos(2\pi f_g t))$ \\
& Hip & $-\frac{\pi}{2}(1 - \cos(2\pi f_g t))$ \\
\hline
\multirow{4}{*}{$S5$: Sitting Down}
& Torso, Head, Shoulders, Elbows, Hands, Ankles & $0$ \\
& Hip & $-\frac{\pi}{2} p(t)$ \\
& Right Knee & $\frac{\pi}{2} p(t)$ \\
& Left Knee & $\frac{\pi}{2} p(t)$ \\
\hline
\multirow{4}{*}{$S6$: Standing Up}
& Torso, Head, Shoulders, Elbows, Hands, Ankles & $0$ \\
& Hip & $-\frac{\pi}{2} (1 - p(t))$ \\
& Right Knee & $\frac{\pi}{2} (1 - p(t))$ \\
& Left Knee & $\frac{\pi}{2} (1 - p(t))$ \\
\hline
$S7$: Body Rotating
& Torso, Head, Shoulders, Elbows, Hands, Hip, Knees, Ankles & $0$ (Modeled in $\mathbf{R}_\mathrm{global}(t)$) \\
\hline
\multirow{4}{*}{$S8$: Walking} 
& Torso, Head, Shoulders, Hands, Hip & $0$ \\
& Right \& Left Elbows & $\pm A_{\text{arm}}\sin(2\pi f_g t + \pi)$ \\
& Right \& Left Knees & $\pm A_{\text{thigh}}\sin(2\pi f_g t)$ \\
& Right \& Left Ankles & $\pm A_{\text{calf}}\sin(2\pi f_g t + \pi/4)$ \\
\hline
\multirow{9}{*}{$S9$: Sitting to Walking}
& Torso, Head, Shoulders, Hands, Hip & $0$ \\
& Right \& Left Elbows & $\begin{cases} 0 & t < t_w \\ \pm A_{\text{arm}}\sin(2\pi f_g (t-t_w)+\pi) & t \ge t_w \end{cases}$ \\
& Right \& Left Knees & $\begin{cases} -\frac{\pi}{2} & t < t_w \\ \pm A_{\text{thigh}}\sin(2\pi f_g (t-t_w)) & t \ge t_w \end{cases}$ \\
& Right \& Left Ankles & $\begin{cases} 0 & t < t_w \\ \pm A_{\text{calf}}\sin(2\pi f_g (t-t_w)+\pi/4) & t \ge t_w \end{cases}$ \\
\hline
\multirow{9}{*}{$S10$: Walking to Sitting}
& Torso, Head, Shoulders, Hands, Hip & $0$ \\
& Right \& Left Elbows & $\begin{cases} \pm A_{\text{arm}}\sin(2\pi f_g t+\pi) & t < t_s \\ 0 & t \ge t_s \end{cases}$ \\
& Right \& Left Knees & $\begin{cases} \pm A_{\text{thigh}}\sin(2\pi f_g t) & t < t_s \\ -\frac{\pi}{2} p(t-t_s) & t \ge t_s \end{cases}$ \\
& Right \& Left Ankles & $\begin{cases} \pm A_{\text{calf}}\sin(2\pi f_g t+\pi/4) & t < t_s \\ 0 & t \ge t_s \end{cases}$ \\
\hline
\multirow{9}{*}{$S11$: Falling to Walking}
& Torso, Head, Shoulders, Hands, Hip & $0$ \\
& Right \& Left Elbows & $\begin{cases} 0 & t < t_w \\ \pm A_{\text{arm}}\sin(2\pi f_g (t-t_w)+\pi) & t \ge t_w \end{cases}$ \\
& Right \& Left Knees & $\begin{cases} 0 & t < t_w \\ \pm A_{\text{thigh}}\sin(2\pi f_g (t-t_w)) & t \ge t_w \end{cases}$ \\
& Right \& Left Ankles & $\begin{cases} 0 & t < t_w \\ \pm A_{\text{calf}}\sin(2\pi f_g (t-t_w)+\pi/4) & t \ge t_w \end{cases}$ \\
\hline
\multirow{9}{*}{$S12$: Walking to Falling}
& Torso, Head, Shoulders, Hands, Hip & $0$ \\
& Right \& Left Elbows & $\begin{cases} \pm A_{\text{arm}}\sin(2\pi f_g t+\pi) & t < t_f \\ 0 & t \ge t_f \end{cases}$ \\
& Right \& Left Knees & $\begin{cases} \pm A_{\text{thigh}}\sin(2\pi f_g t) & t < t_f \\ 0 & t \ge t_f \end{cases}$ \\
& Right \& Left Ankles & $\begin{cases} \pm A_{\text{calf}}\sin(2\pi f_g t+\pi/4) & t < t_f \\ 0 & t \ge t_f \end{cases}$ \\
\hline\hline
\end{tabular}
}
\end{center}
\footnotesize $^{*}$ All variables are consistent with the theoretical derivation. In the code implementation, the starting time for the $S9$ and $S11$ is set to a non-zero constant.\\
\vspace{-0.4cm}
\end{table*}\par

\textbf{$\mathbf{S12}$ - Walking to Falling:}\par
In this activity, the human is walking and then suddenly falls forward. A two-phase activity is modeled. The falling is modeled as a rapid, uncontrolled whole-body rotation to the ground. Let the fall begin at $t_{f}$.\par
Phase 1: Walking ($0 \le t < t_{f}$): Standard $S8$ model is introduced.\par
Phase 2: Falling ($t \ge t_{f}$):  The final walking position of the torso is first recorded. Then, $P(t')$, where $t'=t-t_{f}$, is used to govern the fall dynamics. The body's global rotation angle increases from $0$ to $\pi/2$, and the torso's height decreases to around $0.15~m$:

\vspace{-0.2cm}
\begin{equation}
\theta_{\mathrm{global}}(t) = \frac{\pi}{2} \cdot P(t'),
\end{equation}

\vspace{-0.2cm}
\begin{equation}
z_1(t) = h_{\mathrm{torso}} (1 - P(t')) + 0.15\cdot P(t').
\end{equation}\par
Additionally, a fixed-proportion modeling approach for individuals of varying heights and weights is employed. Distances between joints are scaled proportionally based on height, while the RCS of each joint is scaled proportionally by taking the square root of the product of height and weight. The joint connection rules, distances, and RCS modeling are shown in Table \ref{Human Body Model}. The rotational functions for each joint under various activities are shown in Table \ref{Rotation Functions}.\par

\subsection{Radar Echo Model}
The FMCW waveform is employed for transmission and reception. The instantaneous frequency $f_{\mathrm{TX}}(\hat{t})$ linearly sweeps with a bandwidth of $B$ and a pulse duration of $T_p$. The instantaneous frequency at any fast-time $\hat{t} \in [0, T_p]$ is given by the linear equation:

\vspace{-0.2cm}
\begin{equation}
f_{\mathrm{TX}}(\hat{t}) = f_c + K \hat{t},
\end{equation}
where $f_c$ is the starting carrier frequency and $K = B/T_p$ is the chirp rate. The instantaneous angular frequency is $\omega_{\mathrm{TX}}(\hat{t}) = 2\pi f_{\mathrm{TX}}(\hat{t})$. The phase of the signal $\varphi_{\mathrm{TX}}(\hat{t})$ is the definite integral of the instantaneous angular frequency from the start of the pulse to time $\hat{t}$:

\vspace{-0.2cm}
\begin{equation}
\begin{aligned}
\varphi_{\mathrm{TX}}(\hat{t}) & = \int_0^{\hat{t}} \omega_{\mathrm{TX}}(t') dt' \\&= \int_0^{\hat{t}} (2\pi f_c + 2\pi K t') dt' \\& = [2\pi f_c t' + \pi K (t')^2]_0^{\hat{t}} \\& = 2\pi f_c \hat{t} + \pi K \hat{t}^2
\end{aligned}.
\end{equation}\par
The complex representation of the transmitted signal is:

\vspace{-0.2cm}
\begin{equation}
s_{\mathrm{TX}}(\hat{t}) = e^{j\varphi_{\mathrm{TX}}(\hat{t})}.
\end{equation}\par
The electromagnetic wave propagates from the transmitter location $\mathbf{p}_{\mathrm{TX}}$ to the $i$-th scatterer location $\mathbf{p}_i(t)$ and reflects toward the receiver location $\mathbf{p}_{\mathrm{RX}}$. The total path length at slow-time $t$ is the sum of the transmit and receive path lengths:

\vspace{-0.2cm}
\begin{equation}
R_i(t) = \|\mathbf{p}_i(t) - \mathbf{p}_{\mathrm{TX}}\| + \|\mathbf{p}_i(t) - \mathbf{p}_{\mathrm{RX}}\|.
\end{equation}\par
The round-trip propagation delay is $\tau_i(t) = R_i(t) / c$, where $c$ is the speed of light. The signal received from scatterer $i$ at a specific fast-time $\hat{t}$ is an attenuated replica of the signal that was transmitted at the earlier time $\hat{t} - \tau_i(t)$. Therefore, the phase of the received signal is the phase of the transmitted signal evaluated at this delayed time:

\vspace{-0.2cm}
\begin{equation}
\varphi_{\mathrm{RX}, i}(\hat{t}) = \varphi_{\mathrm{TX}}(\hat{t} - \tau_i(t)).
\end{equation}\par
The complex received signal from this single scatterer is expressed as:

\vspace{-0.2cm}
\begin{equation}
s_{\mathrm{RX}, i}(\hat{t}) = A_i(t) e^{j\varphi_{\mathrm{RX}, i}(\hat{t})},
\end{equation}
where the complex amplitude $A_i(t)$ accounts for path loss and the scatterer's RCS.\par
In the receiver, the incoming signal is mixed with a reference signal, which is the complex conjugate of the transmitted signal, $s_{\mathrm{ref}}^*(\hat{t}) = s_{\mathrm{TX}}^*(\hat{t}) = e^{-j\varphi_{\mathrm{TX}}(\hat{t})}$. This mixing operation, known as de-chirping, is a multiplication in the time domain. The output of the mixer for the signal from scatterer $i$ is the beat signal:

\vspace{-0.2cm}
\begin{equation}
\begin{aligned}
s_{\mathrm{b}, i}(\hat{t}) &= s_{\mathrm{RX}, i}(\hat{t}) \cdot s_{\mathrm{ref}}^*(\hat{t}) \\&= A_i(t) e^{j\varphi_{\mathrm{RX}, i}(\hat{t})} e^{-j\varphi_{\mathrm{TX}}(\hat{t})} \\&= A_i(t) e^{j(\varphi_{\mathrm{RX}, i}(\hat{t}) - \varphi_{\mathrm{TX}}(\hat{t}))}
\end{aligned}.
\end{equation}\par
The phase of this beat signal, $\varphi_{\mathrm{b}, i}(\hat{t})$, is thus the difference between the received and reference phases:

\vspace{-0.2cm}
\begin{equation}
\varphi_{\mathrm{b}, i}(\hat{t}) = \varphi_{\mathrm{TX}}(\hat{t} - \tau_i(t)) - \varphi_{\mathrm{TX}}(\hat{t}).
\end{equation}\par
The full expression for the phase, derived in human kinematic modeling, is substituted:

\vspace{-0.2cm}
\begin{equation}
\begin{aligned}
\varphi_{\mathrm{b}, i}(\hat{t}) &= \left[ 2\pi f_c (\hat{t}-\tau_i(t)) + \pi K(\hat{t}-\tau_i(t))^2 \right] \\&- \left[ 2\pi f_c \hat{t} + \pi K \hat{t}^2 \right].
\end{aligned}
\end{equation}\par
The quadratic term is expanded: $(\hat{t}-\tau_i(t))^2 = \hat{t}^2 - 2\hat{t}\tau_i(t) + \tau_i(t)^2$. This is substituted back into the phase equation:

\vspace{-0.2cm}
\begin{equation}
\begin{aligned}
\varphi_{\mathrm{b}, i}(\hat{t}) &= \left[ 2\pi f_c \hat{t} - 2\pi f_c \tau_i(t) + \pi K (\hat{t}^2 - 2\hat{t}\tau_i(t) + \tau_i(t)^2) \right] \\&- 2\pi f_c \hat{t} - \pi K \hat{t}^2 \\& = (2\pi f_c \hat{t} - 2\pi f_c \hat{t}) + (\pi K \hat{t}^2 - \pi K \hat{t}^2) \\&- 2\pi f_c \tau_i(t) - 2\pi K\hat{t}\tau_i(t) + \pi K \tau_i(t)^2\\&= -2\pi K\tau_i(t)\hat{t} -2\pi f_c \tau_i(t) + \pi K \tau_i(t)^2 \\& \approx -2\pi \left(f_c \tau_i(t) + K \tau_i(t) \hat{t}\right)
\end{aligned}.
\end{equation}\par
The continuous beat signal is sampled at a rate of $f_s$. The continuous fast-time variable is thus replaced by its discrete counterpart, $\hat{t} = n T_s$, where $n$ is the sample index and $T_s=1/f_s$. The slow-time is discretized into pulses, $t = t_m = m T_{\mathrm{PRI}}$, where $m$ is the pulse index and $T_{\mathrm{PRI}} = 1/\mathrm{PRF}$. The total received signal is the coherent sum of contributions from all $N$ scatterers. This yields the discrete raw data matrix:

\vspace{-0.2cm}
\begin{equation}
s_{\mathrm{b}}[n, m] = \sum_{i=1}^{N} A_i(t_m) e^{-j2\pi \left(f_c \tau_i(t_m) + K \tau_i(t_m) n T_s \right)}.
\end{equation}\par
Finally, to simulate a realistic measurement, AWGN $w[n,m]$ is introduced, which is a complex random variable where both the real and imaginary parts are drawn from a Gaussian distribution with zero mean and variance $\sigma_n^2/2$. The final simulated echo is:

\vspace{-0.2cm}
\begin{equation}
s_{\mathrm{noisy}}[n, m] = s_{\mathrm{b}}[n, m] + w[n, m].
\end{equation}\par

\section{Signal Processing Method}
\begin{figure}
    \centering
    \includegraphics[width=0.48\textwidth]{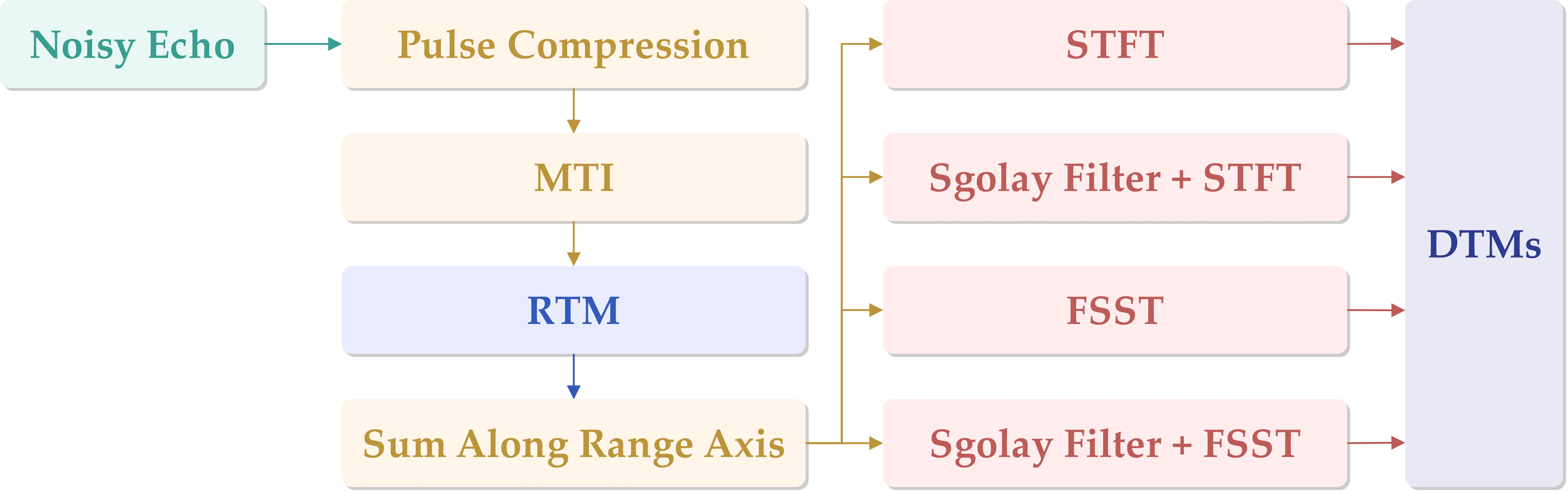}
    \caption{Signal processing overflow.}
    \label{Signal_Processing_Chain}
    \vspace{-0.3cm}
\end{figure}

As shown in Fig. \ref{Signal_Processing_Chain}, the generation methods for RTM and various DTMs are introduced in this section.\par

\subsection{RTM Generation}
The moving target indication (MTI) filter is applied along the slow-time with pulse index $m$ to suppress this static clutter:

\vspace{-0.2cm}
\begin{equation}
s_{\mathrm{MTI}}[n, m] = s_{\mathrm{noisy}}[n, m+1] - s_{\mathrm{noisy}}[n, m].
\end{equation}\par
The beat frequency for each scatterer $i$ is $f_{\mathrm{b},i} = K\tau_i$. The range $R_i$ is related to the delay by $R_i = c\tau_i/2$. Combining these gives a direct linear relationship between range and beat frequency:

\vspace{-0.2cm}
\begin{equation}
f_{\mathrm{b},i} = K \frac{2 R_i}{c} \implies R_i = \frac{c}{2K} f_{\mathrm{b},i}.
\end{equation}\par
The Fast Fourier Transform (FFT) is performed along the fast-time axis with index $n$ for each pulse $m$ of the MTI-filtered data:

\vspace{-0.2cm}
\begin{equation}
S[k, m] = \sum_{n=0}^{N_{\mathrm{ADC}}-1} s_{\mathrm{MTI}}[n, m] e^{{-j\frac{2\pi nk}{N_{\mathrm{FFT}}}}},
\end{equation}
where $k$ is the discrete frequency bin index, $N_{\mathrm{ADC}}$ is the number of sampling points, $N_{\mathrm{FFT}}$ is the number of FFT points. Each bin index $k$ corresponds to a discrete frequency $f_k = k \cdot (f_s/N_{\mathrm{FFT}})$. Substituting this into the range equation provides the mapping from FFT bin index to physical range:

\vspace{-0.2cm}
\begin{equation}
R_k = \frac{c}{2K} f_k = \frac{c f_s}{2 K N_{\mathrm{FFT}}} k.
\end{equation}\par
The magnitude of the resulting complex matrix $|S[k, m]|$, which represents the signal strength at each range bin $k$ for each pulse $m$, is the desired RTM.\par

\subsection{DTMs Generation}
For Doppler analysis, the phase information across time is paramount. A single complex time-series representing the aggregate motion is created by coherently summing the complex values in the range-processed matrix $S[k,m]$ along the range axis with index $k$:

\vspace{-0.2cm}
\begin{equation}
x[m] = \sum_{k=0}^{N_{\mathrm{FFT}}/2-1} S[k, m].
\end{equation}\par
For the torso moving with instantaneous radial velocity $v_r(t)$, this is $f_d(t) = -2v_r(t)/\lambda$. The instantaneous phase change is $d\phi(t) = 2\pi f_d(t) dt$. The total phase accumulated due to this bulk motion up to time $t_m$ is the integral:

\vspace{-0.2cm}
\begin{equation}
\phi_{\mathrm{bulk}}(t_m) = \int_0^{t_m} 2\pi f_d(t') dt'.
\end{equation}\par
\par
In the discrete-time domain, where the Doppler frequency is calculated for each pulse $i$, this integral is approximated by a cumulative sum:

\vspace{-0.2cm}
\begin{equation}
\phi_{\mathrm{bulk}}[m] = \sum_{i=1}^{m} 2\pi f_{d,\text{bulk}}[i] \cdot T_{\mathrm{PRI}},
\end{equation}
where $f_{d,\text{bulk}}[i]$ represents the bulk Doppler shift. To remove it, the signal $x[m]$ is multiplied by a complex phasor that rotates in the opposite direction. This demodulation operation yields the compensated signal:

\vspace{-0.2cm}
\begin{equation}
x_{\mathrm{comp}}[m] = x[m] \cdot e^{-j\phi_{\mathrm{bulk}}[m]}.
\end{equation}\par
For measured data, it is recommended to detect the torso joint via RTM, estimate its velocity and Doppler shift, and then implement compensation. After bulk Doppler compensation, the complex slow-time signal is denoised using a Savitzky-Golay filter \cite{Sgolay}. For each data point, a window of adjacent points is selected, and a low-degree polynomial is fitted to this window using the method of linear least squares. Let the signal be $x[m] = x_{\mathrm{Re}}[m] + j \cdot x_{\mathrm{Im}}[m]$. The process is:

\vspace{-0.2cm}
\begin{equation}
\begin{gathered}
x_{\mathrm{Re, denoised}}[m] = \text{SG}[x_{\mathrm{Re}}[m], P, F]\\
x_{\mathrm{Im, denoised}}[m] = \text{SG}[x_{\mathrm{Im}}[m], P, F]\\
x_{\mathrm{denoised}}[m] = x_{\mathrm{Re, denoised}}[m] + j \cdot x_{\mathrm{Im, denoised}}[m]
\end{gathered},
\end{equation}
where $P$ is the polynomial order and $F$ is the frame length, $\text{SG}[\cdot]$ is the Savitzky-Golay filter.\par
The final step is a time-frequency representation of the fully processed slow-time signal $x_{\mathrm{denoised}}[m]$. The STFT is employed to analyze how the Doppler frequency content of the signal evolves over time. The signal is segmented by a sliding window function $g[\cdot]$, and an FFT is computed for each segment:

\vspace{-0.2cm}
\begin{equation}
X_{\mathrm{STFT}}[l, \omega] = \sum_{m=-\infty}^{\infty} x_{\mathrm{denoised}}[m] g[m-l \cdot H] e^{-j\omega m},
\end{equation}
where $l$ is the discrete time shift index and $H$ is the hop size. The resulting spectrogram $|X_{\mathrm{STFT}}[l, \omega]|^2$ is the desired DTM.\par
To achieve a higher-resolution DTM, the FSST is also employed. This is a post-processing technique applied to the STFT result. For each point $(l, \omega)$ in the STFT plane where there is significant energy, an instantaneous frequency estimation $\omega_\mathrm{if}(l, \omega)$ is calculated. The energy of the STFT at point $(l, \omega)$ is squeezed along the frequency axis to the new location $(l, \omega_\mathrm{if})$, which results in a significantly sharpened and more interpretable DTM.\par

\section{FFT-Based Neural Network Design}
\begin{figure}[!ht]
    \centering
    \includegraphics[width=0.48\textwidth]{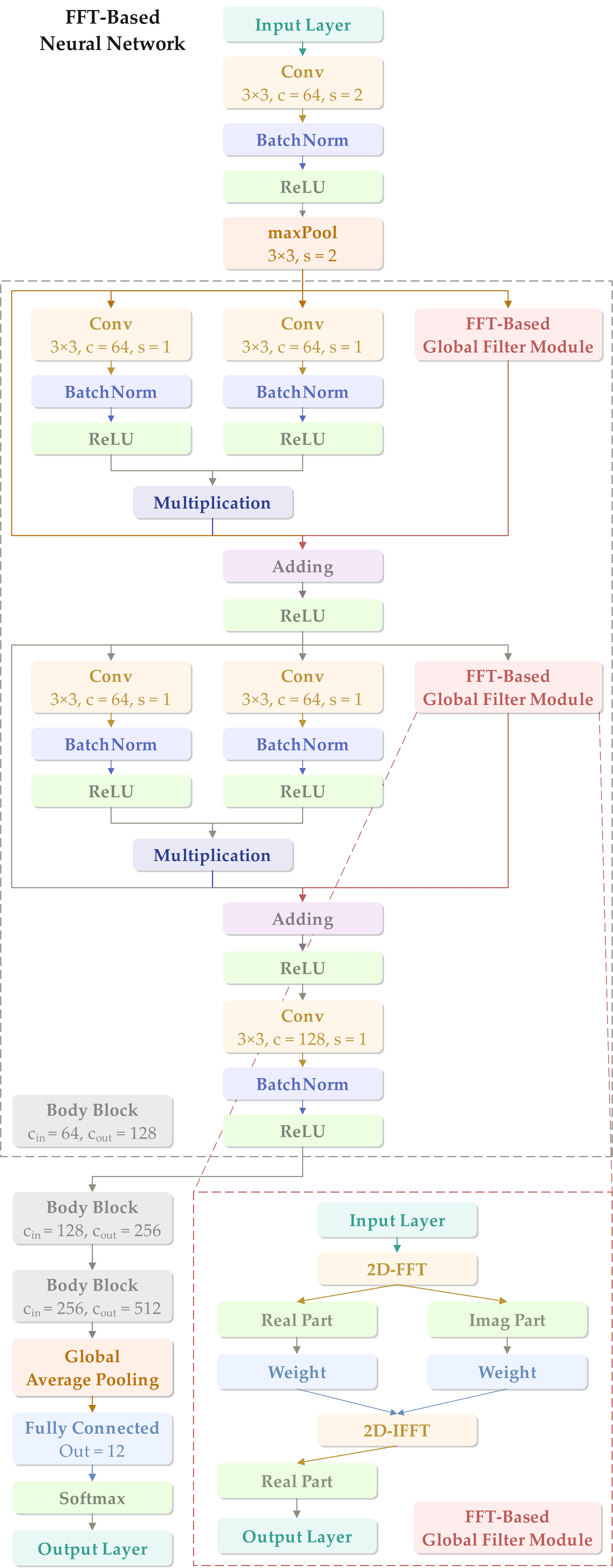}
    \caption{Structure of the proposed FFT-based neural network model.}
    \label{FFTBasedNN}
    \vspace{-0.3cm}
\end{figure}

For the task of radar-based HAR from DTMs, a novel neural network architecture is designed shown in Fig. \ref{FFTBasedNN}. The core principle of this design is the replacement of the computationally intensive self-attention mechanisms commonly found in ViT with FFT-based global filter module. This module is engineered to facilitate the exchange of information across all spatial locations of the feature maps by operating in the frequency domain. The architecture is constructed to improve training stability and lightweight performance \cite{StarNet}.\par

\subsection{FFT-Based Global Filter Module}
The FFT-based global filter module is achieved through three sequential operations \cite{GFNet}. First, an input feature map from the spatial domain, denoted as $\boldsymbol{x}$, is transformed into the frequency domain using a two-dimensional (2D) FFT. This transformation is expressed by the following equation:

\vspace{-0.2cm}
\begin{equation}
\mathbf{X}_\mathrm{GF}=\mathrm{FFT_\mathrm{2D}}[\mathbf{x}_\mathrm{GF}] \in \mathbb{C}^{H_\mathrm{GF} \times W_\mathrm{GF} \times D_\mathrm{GF}},
\end{equation}
where $H_\mathrm{GF}, W_\mathrm{GF}, D_\mathrm{GF}$ represent the height, width, and channel depth of the feature map, respectively.\par
Second, in the frequency domain, the resulting complex-valued feature map $\mathbf{X}_\mathrm{GF}$ is subjected to an element-wise multiplication with a set of learnable global filters, $\mathbf{K}_\mathrm{GF}$. This operation is defined as:

\vspace{-0.2cm}
\begin{equation}
\mathbf{X}_\mathrm{GF}^{\prime}=\mathbf{K}_\mathrm{GF} \odot \mathbf{X}_\mathrm{GF}.
\end{equation}\par
The filter $\mathbf{K}_\mathrm{GF}$ is a complex-valued tensor of the same dimensions as $\mathbf{X}_\mathrm{GF}$ and its parameters are learned during the training process. For implementation, these complex weights are stored as a real-valued array of size $[H_\mathrm{GF}, W_\mathrm{GF}, D_\mathrm{GF}, 2]$, where the last dimension separates the real and imaginary components.\par
Finally, the filtered feature map $\tilde{\mathbf{X}_\mathrm{GF}}$ is converted back to the spatial domain using a 2D inverse FFT. The real part of the resulting tensor is then passed to the subsequent layer in the network. The transformation is given by:

\vspace{-0.2cm}
\begin{equation}
x_\mathrm{GF} \leftarrow \mathrm{FFT_\mathrm{2D}}^{-1}[\mathbf{X}_\mathrm{GF}^{\prime}].
\end{equation}\par
Due to the use of complex numbers and FFT operations, for which gradients can be complex-valued, a custom backward pass function is implemented for this module. This ensures that real-valued gradients are returned to the optimizer, allowing for stable training with standard optimizers like Adam.\par

\subsection{Network Architecture}
The network accepts any one of eight generated DTMs with input dimensions of $256 \times 256 \times 3$. The initial processing stage, or stem, consists of a $3 \times 3$ convolutional layer with a stride of $2$, followed by a batch normalization layer, a ReLU activation function, and a $3 \times 3$ max-pooling layer, also with a stride of $2$. This stem is responsible for initial feature extraction and reduces the spatial dimensions of the input to $64 \times 64$.\par
\par
The main body of the network is composed of a series of blocks, with downsampling layers placed between stages to progressively reduce spatial resolution and increase channel depth. The channel dimensions are sequentially increased from $64$, to $128$, to $256$, and finally to $512$. Each block in the body contains three parallel branches that are later combined: (1) Convolutional Branches: Two parallel paths, each containing a $3 \times 3$ convolution, batch normalization, and a ReLU activation. The outputs of these two paths are combined via an element-wise multiplication layer. (2) Global Filter Branch: A third parallel path which consists of the FFT-based global filter module described previously. (3) Residual Connection: A skip connection is made from the input of the block. The outputs of the multiplied convolutional branches, the global filter module, and the residual connection are all summed together using an addition layer. This is followed by a final ReLU activation. This block structure is repeated, allowing the network to learn both local features through convolutions and global relationships through the FFT-based filtering. \par
Following the final body block, a global average pooling layer is used to reduce each feature map to a single value. This is followed by a fully connected layer with $12$ output units, corresponding to the $12$ classes of human activities. A softmax layer is then applied to produce a probability distribution over the classes, and a final classification layer computes the loss.\par
For regularization and to improve model generalization, the network is trained using a label smoothing cross-entropy loss function \cite{Label-Smoothing}. The label smoothing cross-entropy loss for a single sample is formulated as follows:

\vspace{-0.2cm}
\begin{equation}
L_\mathrm{LS} = -\sum_{k_\mathrm{Cls}=1}^{K_\mathrm{Cls}} y^{\prime}_{\mathrm{cls},k} \log(p_{\mathrm{cls},k}),
\end{equation}
where $K_\mathrm{Cls}$ is the total number of classes, $p_{\mathrm{cls},k}$ is the predicted probability for class $k_\mathrm{Cls}$, and $y^{\prime}_{\mathrm{cls},k}$ is the smoothed label for class $k_\mathrm{Cls}$. The smoothed label $y^{\prime}_{\mathrm{cls},k}$ is calculated from the original one-hot label $y_{\mathrm{cls},k}$ using the formula:

\vspace{-0.2cm}
\begin{equation}
y^{\prime}_{\mathrm{cls},k} = y_{\mathrm{cls},k}(1-\zeta) + \frac{\zeta}{K_\mathrm{Cls}}
\end{equation}
where $\zeta$ is a small hyperparameter. This technique discourages the model from assigning the full probability mass to a single class, which can help prevent overfitting and lead to a more calibrated model. The network has a total of $74$ layers and $6.50~\mathrm{M}$ parameters, exhibiting certain lightweight characteristics.\par
\begin{table}
\begin{center}
\caption{Experimental Parameter Settings.\label{tab:SimParams}}
\vspace{-0.0cm}
\resizebox{0.48\textwidth}{!}{
\begin{tabular}{ccc}
\hline\hline
\textbf{Parameter} & \textbf{Free-space} & \textbf{Through-the-wall} \\
\hline
\multicolumn{3}{c}{\textbf{Radar Parameters}} \\
\hline
Carrier Frequency & $77\mathrm{~GHz}$ & $2\mathrm{~GHz}$ \\
Bandwidth & $4\mathrm{~GHz}$ & $1\mathrm{~GHz}$ \\
Pulse Repetition Frequency & $5000\mathrm{~Hz}$ & $128\mathrm{~Hz}$ \\
Sampling Frequency & \multicolumn{2}{c}{$10\mathrm{~MHz}$} \\
Pulse Duration & \multicolumn{2}{c}{$40~\mu s$} \\
Transmitter Position & \multicolumn{2}{c}{$(0.4, -0.1, 1.5)~m$} \\
Receiver Position & \multicolumn{2}{c}{$(0.4, 0.1, 1.5)~m$} \\
Antenna Gain & \multicolumn{2}{c}{$6\mathrm{~dBi}$} \\
Antenna Isolation & \multicolumn{2}{c}{$10\mathrm{~dB}$} \\
SNR & \multicolumn{2}{c}{$30\mathrm{~dB}$} \\
\hline
\multicolumn{3}{c}{\textbf{Kinematic Parameters}} \\
\hline
Height & \multicolumn{2}{c}{$1.8~m$} \\
Weight & \multicolumn{2}{c}{$84.24\mathrm{~kg}$} \\
Initial Position & \multicolumn{2}{c}{$(2,0)~m$ or $(1,0)~m$$^{1}$} \\
Torso Velocity & \multicolumn{2}{c}{$1.5~m/s$} \\
Motion Angle & \multicolumn{2}{c}{$0^{\circ}$} \\
Gait Frequency & \multicolumn{2}{c}{$1\mathrm{~Hz}$} \\
Thigh Rotation Amplitude & \multicolumn{2}{c}{$30^{\circ}$} \\
Calf Rotation Amplitude & \multicolumn{2}{c}{$45^{\circ}$} \\
Arm Rotation Amplitude & \multicolumn{2}{c}{$35^{\circ}$} \\
Simulation Time & \multicolumn{2}{c}{$4~s$} \\
\hline
\multicolumn{3}{c}{\textbf{Wall Parameters}} \\
\hline
Wall Center Position & / & $(1,0,1.25)~m$ \\
Wall Dimensions  & / & $(0.24, 5, 2.5)~m$ \\
Relative Dielectric Constant  & / & $6$ \\
Loss Tangent & / & $0.03$ \\
\hline\hline
\end{tabular}
}
\end{center}
\footnotesize $^{1}$ Initial position is $(2, 0)~m$ for activities $S1 \sim S7$, and $(1, 0)~m$ for activities $S8 \sim S12$.\\
\vspace{-0.2cm}
\end{table}\par
\begin{table}
\begin{center}
\caption{Uniform Hyperparameters for Networks$^{*}$.\label{Training Settings}}
\vspace{-0.1cm}
\resizebox{0.48\textwidth}{!}{
\begin{tabular}{cc}
\hline\hline
\textbf{Name of Hyperparameters}             & \textbf{Value}          \\ \hline
Batch Size                      & $128$                     \\
Total Epoches                   & $20$                    \\
Initial Learning Rate$^{1}$     & $0.00147$                   \\
Regulization Method             & $L - 2$           \\
Optimizer                       & Adam  \\
Solidified Model                & Best Epoch                \\
Training Dataset                & $2880$                     \\
Validation Dataset              & $720$                     \\
Hardware of Training and Validation   & NVIDIA RTX 3060 OC \\
Software of Training and Validation  & Matlab R2025a   \\
\hline\hline
\end{tabular}
}
\end{center}
\vspace{-0.4cm}
\end{table}\par
\begin{figure*}[!ht]
    \centering
    \includegraphics[width=\textwidth]{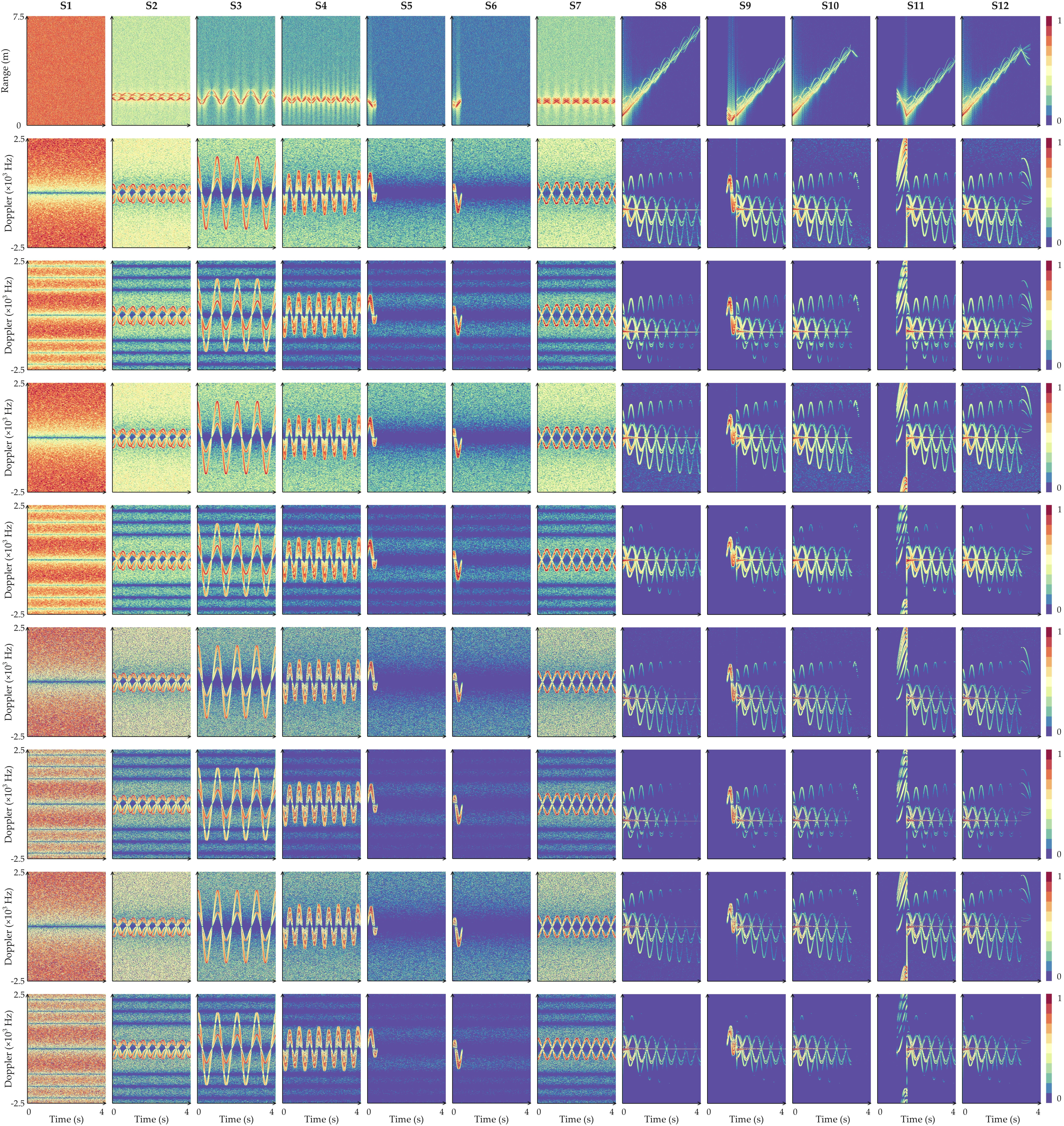}
    \caption{Visualizations for free-space detection. The first line is the RTMs, the second line is the STFT-based DTMs, the third line is the denoised STFT-based DTMs, the fourth line is the compensated STFT-based DTMs, the fifth line is the compensated and denoised STFT-based DTMs, the sixth line is the FSST-based DTMs, the seventh line is the denoised FSST-based DTMs, the eighth line is the compensated FSST-based DTMs, the ninth line is the compensated and denoised FSST-based DTMs.}
    \label{77G_Visualization}
    \vspace{-0.1cm}
\end{figure*}\par
\begin{figure*}[!ht]
    \centering
    \includegraphics[width=\textwidth]{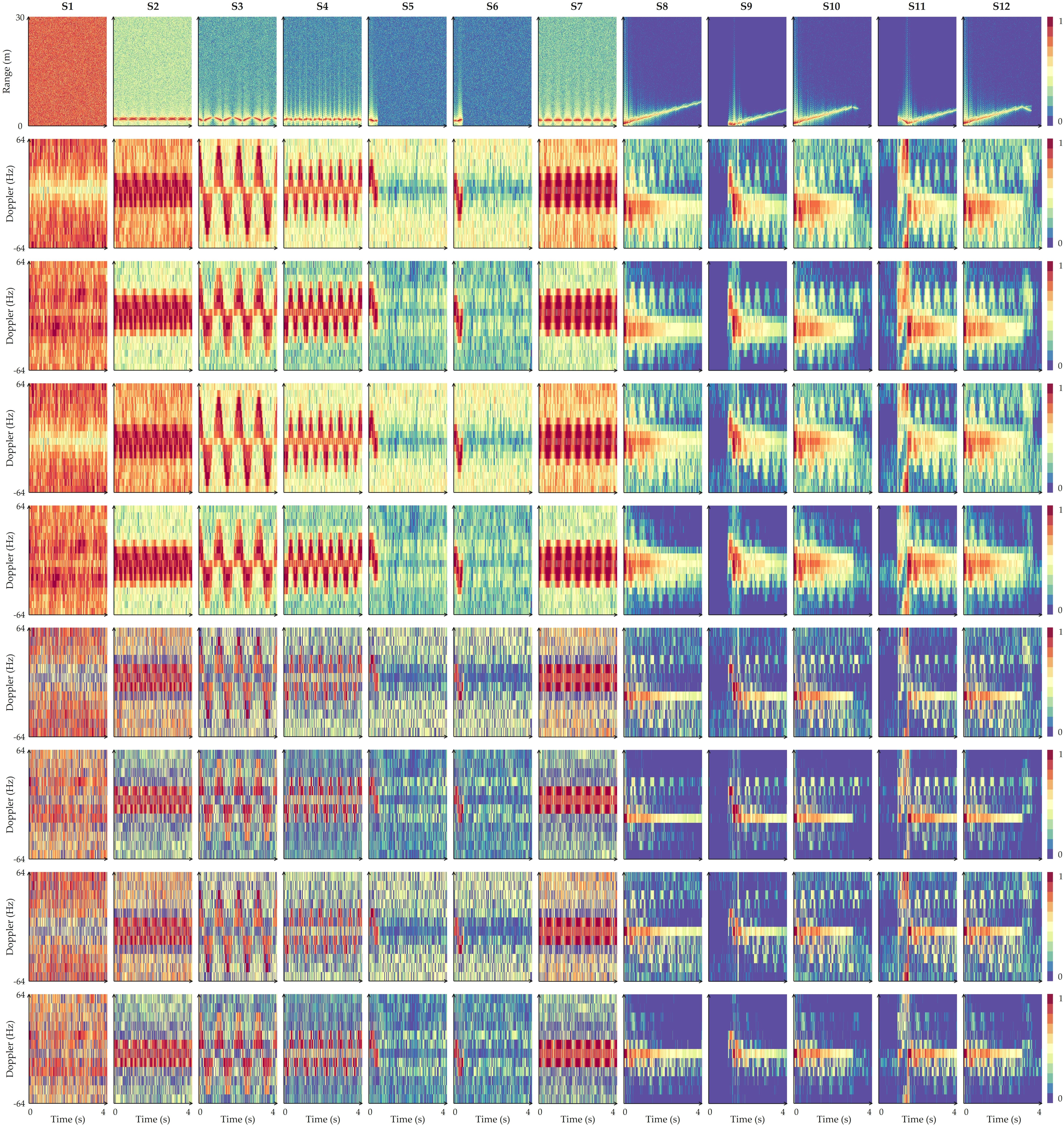}
    \caption{Visualizations for through-the-wall detection. Each row corresponds to the same content shown in Fig. \ref{77G_Visualization}.}
    \label{2G_Visualization}
    \vspace{-0.1cm}
\end{figure*}\par

\section{Numerical Experiments}
The data generated by the designed simulator and the performance of the proposed neural network are comprehensively validated in this section.\par

\subsection{Experimental Settings}
As shown in TABLE \ref{tab:SimParams}, the parameters of both kinematic model and the radar echo model can be adjusted in the simulator. Two scenarios are applied for analysis: Free-space detection \cite{Free-Space Detection} and through-the-wall detection \cite{Through-the-Wall Detection}.\par
In free-space detection scenario, the radar's carrier frequency is set to $77\mathrm{~GHz}$ with a bandwidth of $4\mathrm{~GHz}$. A pulse repetition frequency of $5000\mathrm{~Hz}$ and a sampling frequency of $10\mathrm{~MHz}$ are used, with a pulse duration of $40~\mu s$. The transmitter and receiver are positioned at $(0.4, -0.1, 1.5)~m$ and $(0.4, 0.1, 1.5)~m$, respectively. An antenna gain of $6\mathrm{~dBi}$ and an antenna isolation of $10\mathrm{~dB}$ are specified, and a signal-to-noise ratio (SNR) of $30\mathrm{~dB}$ is assumed. For the kinematic model, the subject is defined with a height of $1.8~m$ and a weight of $84.24\mathrm{~kg}$. The simulation begins with the target at an initial position of $(2, 0)~m$ for activities $S1 \sim S7$, and $(1, 0)~m$ for activities $S8 \sim S12$ moving at a torso velocity of $1.5~m/s$ in a direction of $0$ degrees. The gait frequency is set to $1\mathrm{~Hz}$, while the swing amplitudes of the thigh, calf, and arm are configured to $30$, $45$, and $35$ degrees, respectively. The total simulation time is set to $4$ seconds.\par
In through-the-wall detection scenario, the radar's carrier frequency is changed to $2\mathrm{~GHz}$ with a bandwidth of $1\mathrm{~GHz}$. A pulse repetition frequency of $128\mathrm{~Hz}$ is used. The center of the wall is placed at $(1,0,1.25)~m$. The length, width, and height of the wall are $0.24$, $5$, and $2.5~m$, respectively. The relative dielectric constant and the loss tangent of the wall are $6$ and $0.03$, respectively. The rest of the parameters remain the same as free-space detection.\par

\subsection{Visualizations}
The visualization results for the free-space and through-the-wall detection scenarios are presented. To generate these DTMs, for STFT and FSST, a window with a length corresponding to $10\%$ of the signal duration and a $90\%$ overlap between consecutive windows is utilized. In addition, a Savitzky-Golay filter with a polynomial order of $5$ and a frame length of $11$ is applied to denoise the resulting spectrograms.\par
The results for the free-space detection scenario are displayed in Fig. \ref{77G_Visualization}. The top row illustrates the RTMs, while the subsequent rows present the corresponding DTMs after different stages of processing. For in-place activities ($S1 \sim S7$), the target's range is observed to be constant over time. In the spectrograms, periodic micro-Doppler signature is clearly visible. For dynamic activities that include translational motion ($S8 \sim S12$), a linear change in range over time is evident in the RTM. Correspondingly, the DTMs exhibit a significant bulk Doppler shift associated with the torso's velocity, upon which the periodic micro-Doppler signature from limb movements are superimposed. Similar conclusions can be drawn from Fig. \ref{2G_Visualization} for through-the-wall scenario. In contrast, a noticeable degradation in signal quality is observed when compared to the free-space results. The presence of the wall introduces significant signal attenuation. Consequently, the micro-Doppler signature in the spectrograms appears weaker and is less distinct. The finer details of the limb motions are obscured, and the overall spectrum is contaminated with more background noise. Despite this degradation, the fundamental characteristics that distinguish between in-place ($S1 \sim S7$) and moving ($S8 \sim S12$) targets are still preserved.\par
\begin{figure*}[!ht]
    \centering
    \includegraphics[width=\textwidth]{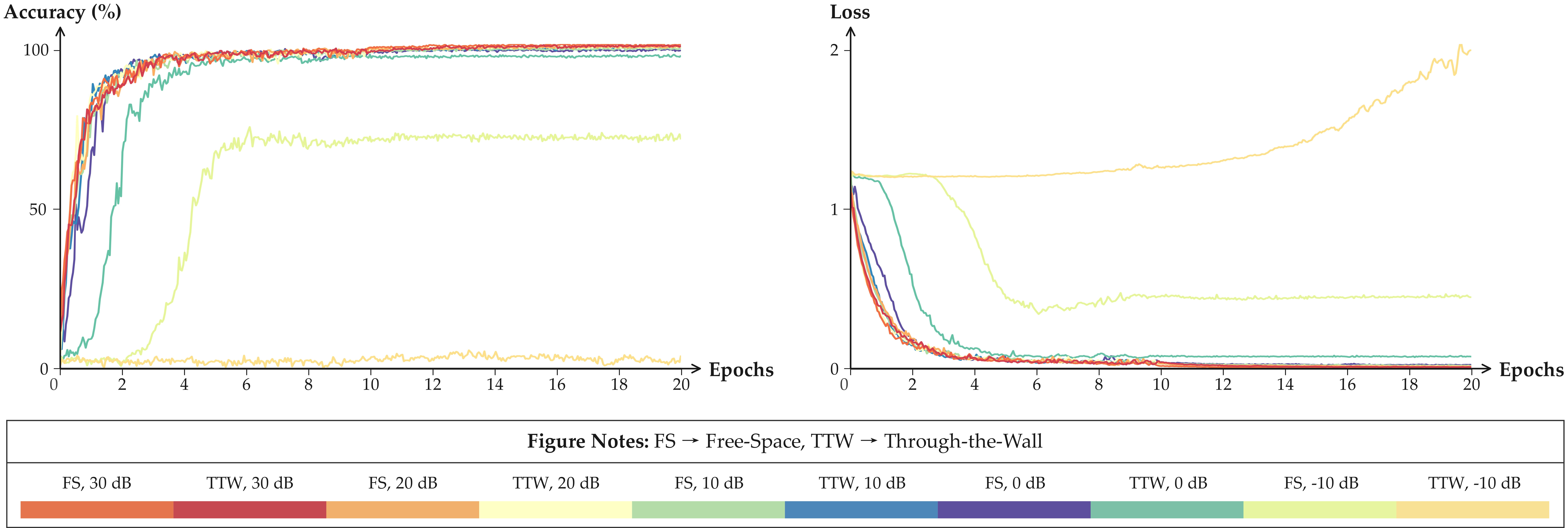}
    \caption{Validation accuracy and loss curves under different scenarios and SNRs.}
    \label{Training_Validation_Curves}
    \vspace{-0.1cm}
\end{figure*}\par
\begin{figure}
    \centering
    \includegraphics[width=0.48\textwidth]{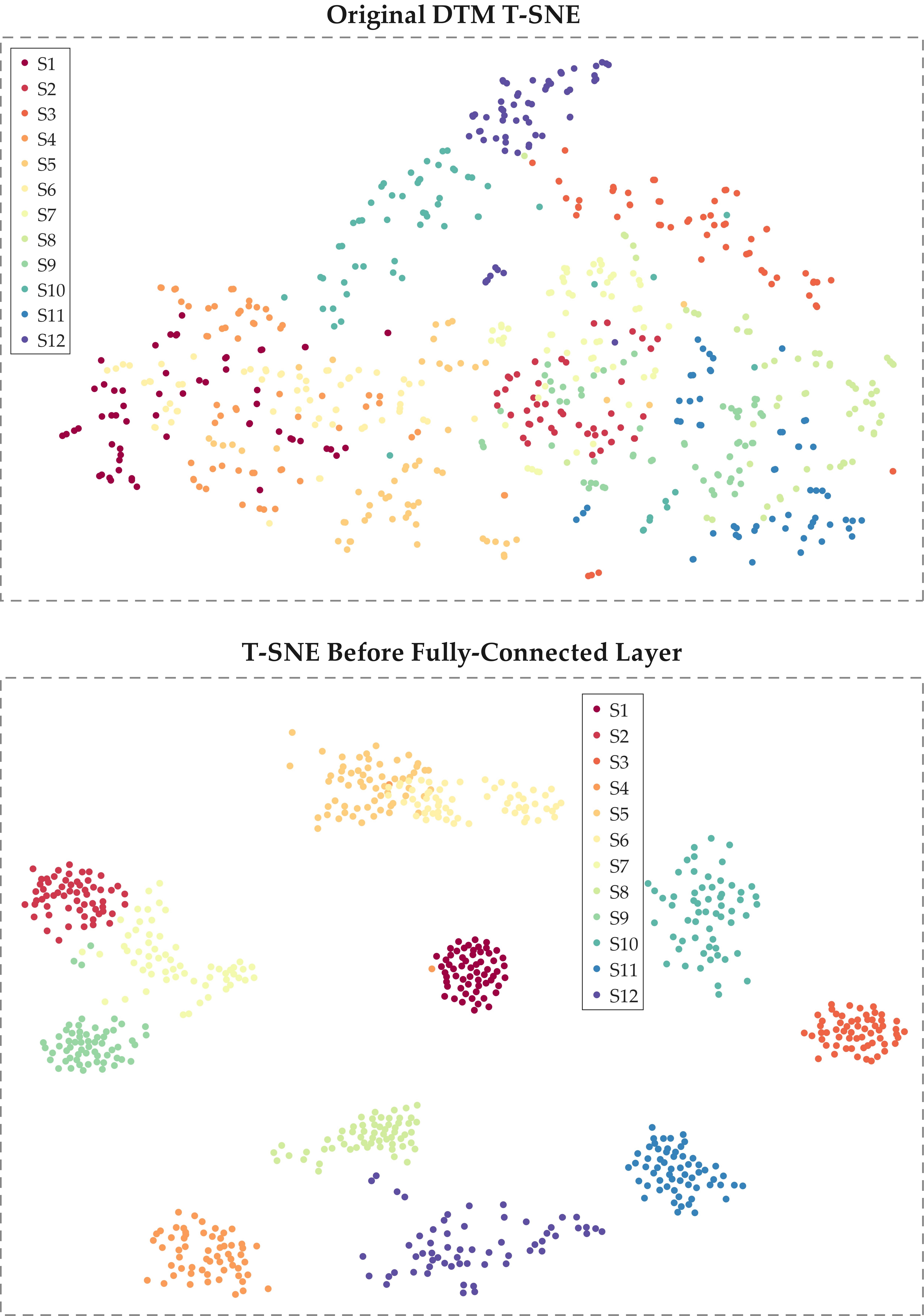}
    \caption{T-SNE results of original STFT-based DTMs and feature maps before the fully-connected layer in free-space scenario.}
    \label{T_SNE_FS}
    \vspace{-0.3cm}
\end{figure}\par
\begin{figure}
    \centering
    \includegraphics[width=0.48\textwidth]{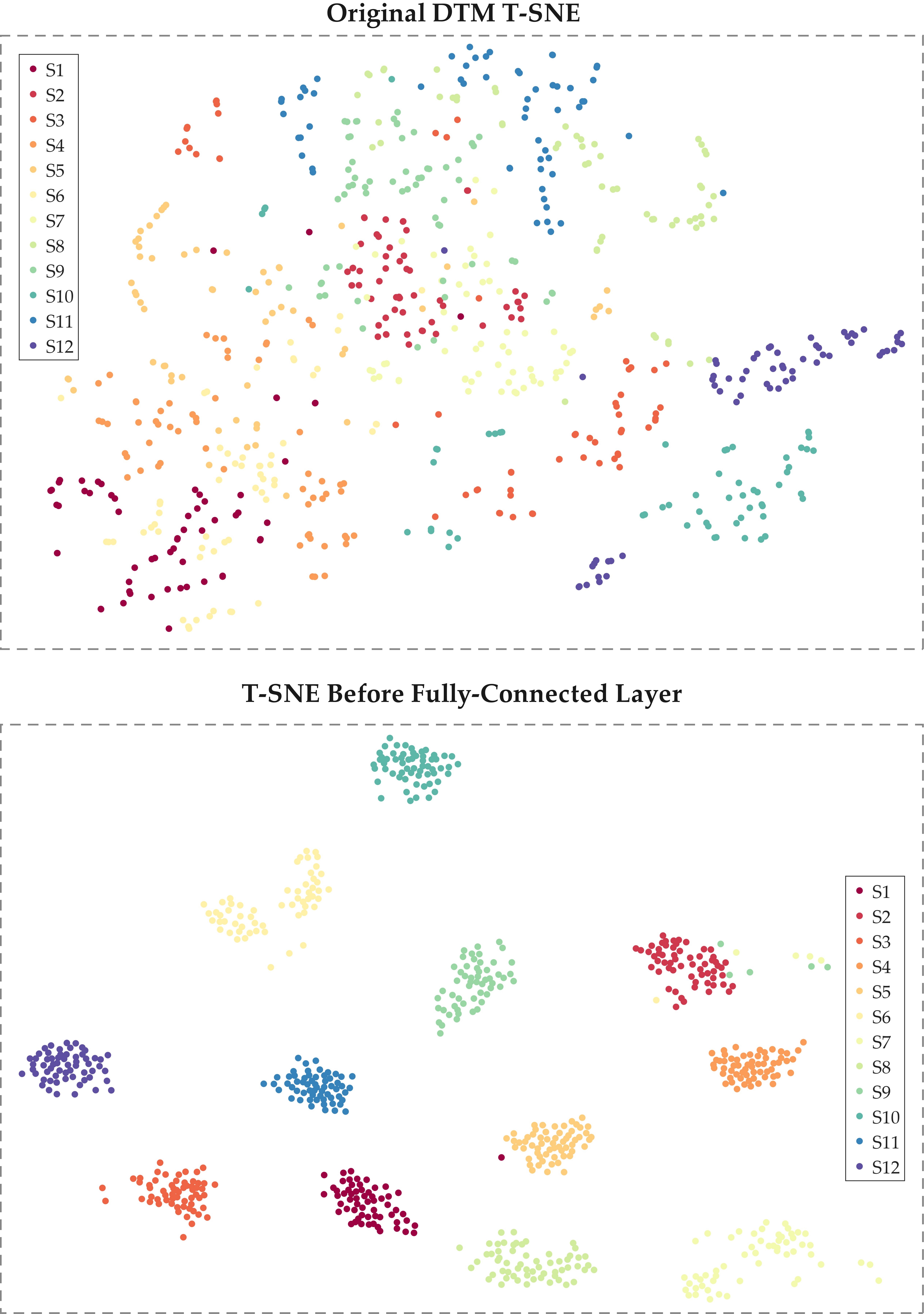}
    \caption{T-SNE results of original STFT-based DTMs and feature maps before the fully-connected layer in through-the-wall scenario.}
    \label{T_SNE_TTW}
    \vspace{-0.3cm}
\end{figure}\par
\begin{figure*}[!ht]
    \centering
    \includegraphics[width=\textwidth]{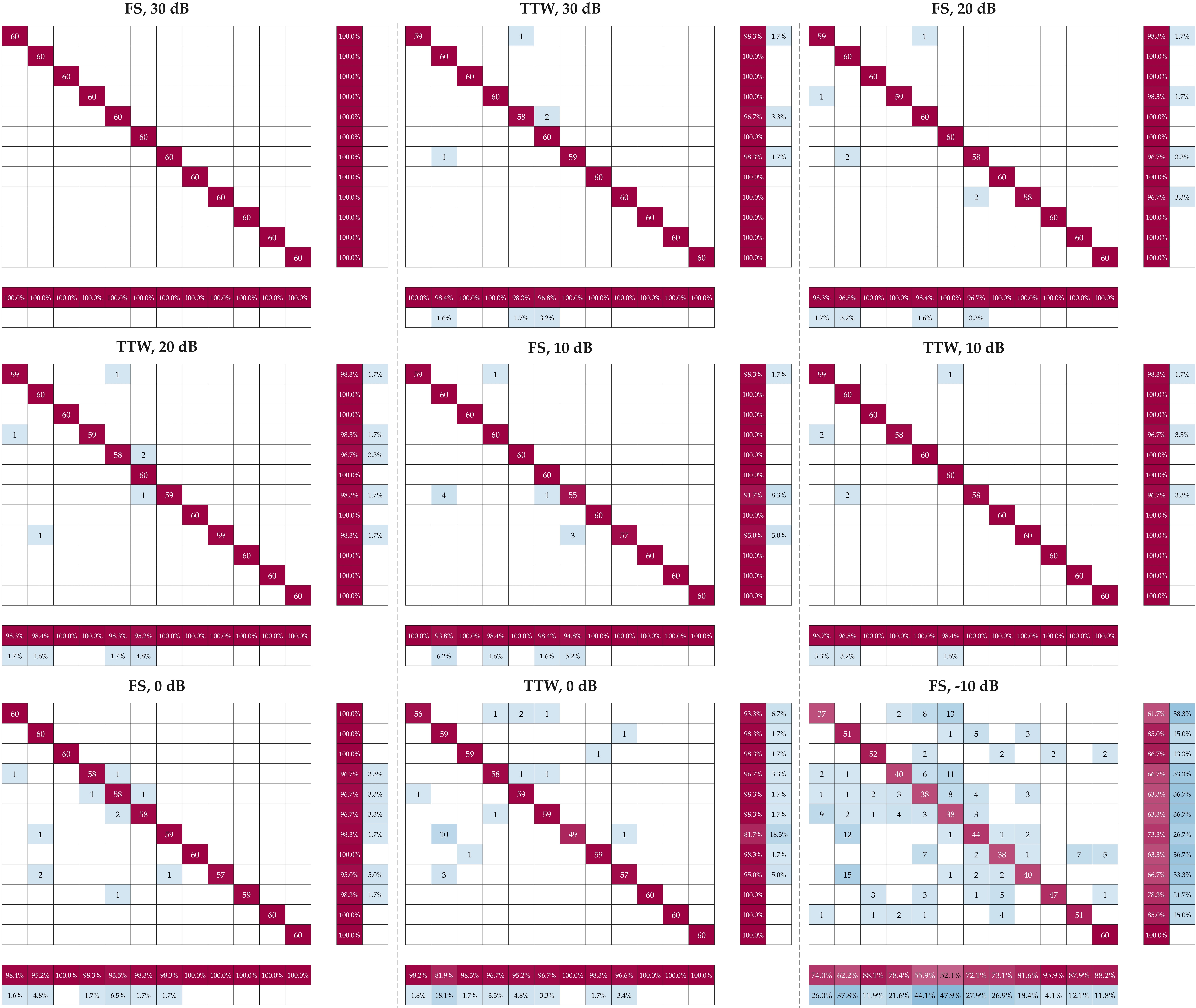}
    \caption{Validation confusion matrices under different scenarios and SNRs.}
    \label{Confusion_Matrics}
    \vspace{-0.3cm}
\end{figure*}\par

\subsection{Accuracy and Robustness Verifications}
The training curves for accuracy and loss on the validation set are presented in Fig. \ref{Training_Validation_Curves}, covering free-space and through-the-wall detection scenarios across SNR levels from $30\mathrm{~dB}$ to $-10\mathrm{~dB}$ over $20$ epochs. The dataset is generated using a designed simulator by producing random values within the following ranges: Human height from $1.5$ to $1.9$ meters, human motion direction angle from $-30$ to $30$ degrees, range coordinate of the human initial position from $1$ to $3$ meters, and rotation amplitude angle of each adjustable human joint within $\pm 20\%$. For each activity, $300$ STFT-based DTMs are generated. These are randomly split into training and validation sets at an $8:2$ ratio. The hyperparameter settings for network training are shown in TABLE \ref{Training Settings}.\par
\begin{table}
\begin{center}
\caption{Accuracy and Cost Comparison Experiments.\label{Comparison}}
\vspace{-0.6cm}
\resizebox{0.5\textwidth}{!}{
\begin{tabular}{cccc}
\hline\hline \textbf{Method} & \textbf{Validation Accuracy ($\%$)}   & \textbf{Parameters (M)} & \textbf{Time (s)}$^{1}$\\
\hline
\multicolumn{4}{c}{\textbf{Free-Space Scenario}}\\
\hline
MobileNet-V2 \cite{MobileNetV2} & $97.50$ & $2.23$ & $0.15$ \\
ResNet-50 \cite{ResNet50} & $98.75$ & $23.58$ & $0.50$ \\
VGG-19 \cite{VGG19} & $99.17$ & $139.55$ & $1.00$ \\
ViT \cite{ViT} & $100.00$ & $85.24$ & $0.80$ \\
ConvNeXt \cite{ConvNeXt} & $100.00$ & $28.24$ & $0.60$ \\
$\textbf{Proposed}$ & $\mathbf{100.00}$ & $\mathbf{6.50}$ & $\mathbf{0.22}$\\
\hline
\multicolumn{4}{c}{\textbf{Through-the-Wall Scenario}}\\
\hline
MobileNet-V2 & $94.86$ & $2.23$ & $0.15$ \\
ResNet-50 & $96.81$ & $23.58$ & $0.50$ \\
VGG-19 & $98.33$ & $139.55$ & $1.00$ \\
ViT & $99.72$ & $85.24$ & $0.80$ \\
ConvNeXt & $99.44$ & $28.24$ & $0.60$ \\
$\textbf{Proposed}$ & $\mathbf{99.44}$ & $\mathbf{6.50}$ & $\mathbf{0.22}$\\
\hline\hline
\end{tabular}
}
\end{center}
\footnotesize $^{1}$ The inference time for one DTM image, including model loading time.\\
\vspace{-0.5cm}
\end{table}
\par
In free-space scenario, accuracy curves for higher SNR rise rapidly to near $100\%$ within initial epochs while loss declines sharply to minima. Lower SNR shows slower ascent stabilizing at $80\sim 90\%$ after $10\sim 15$ epochs with gradual, fluctuating loss reductions due to noise, which demonstrates robust convergence and superior performance from cleaner signals. In contrast, curves under through-the-wall scenario exhibit degradation, with accuracy increasing more slowly and plateauing below $95\%$ even at higher SNR, and loss descending sluggishly with higher residuals from wall attenuation. At lower SNR, prolonged plateaus around $70\sim 85\%$ and volatile loss minimization occur due to attenuation and noise effects, yet fundamental trends of improving accuracy and decreasing loss persist, which highlights the resilience of the proposed model. Unfortunately, for through-the-wall and $\mathrm{SNR}\leq -10\mathrm{~dB}$ scenario, the network model tends to be ineffective. Therefore, subsequent experiments will no longer include results from this scenario for comparison.\par
\begin{figure*}
    \centering
    \includegraphics[width=\textwidth]{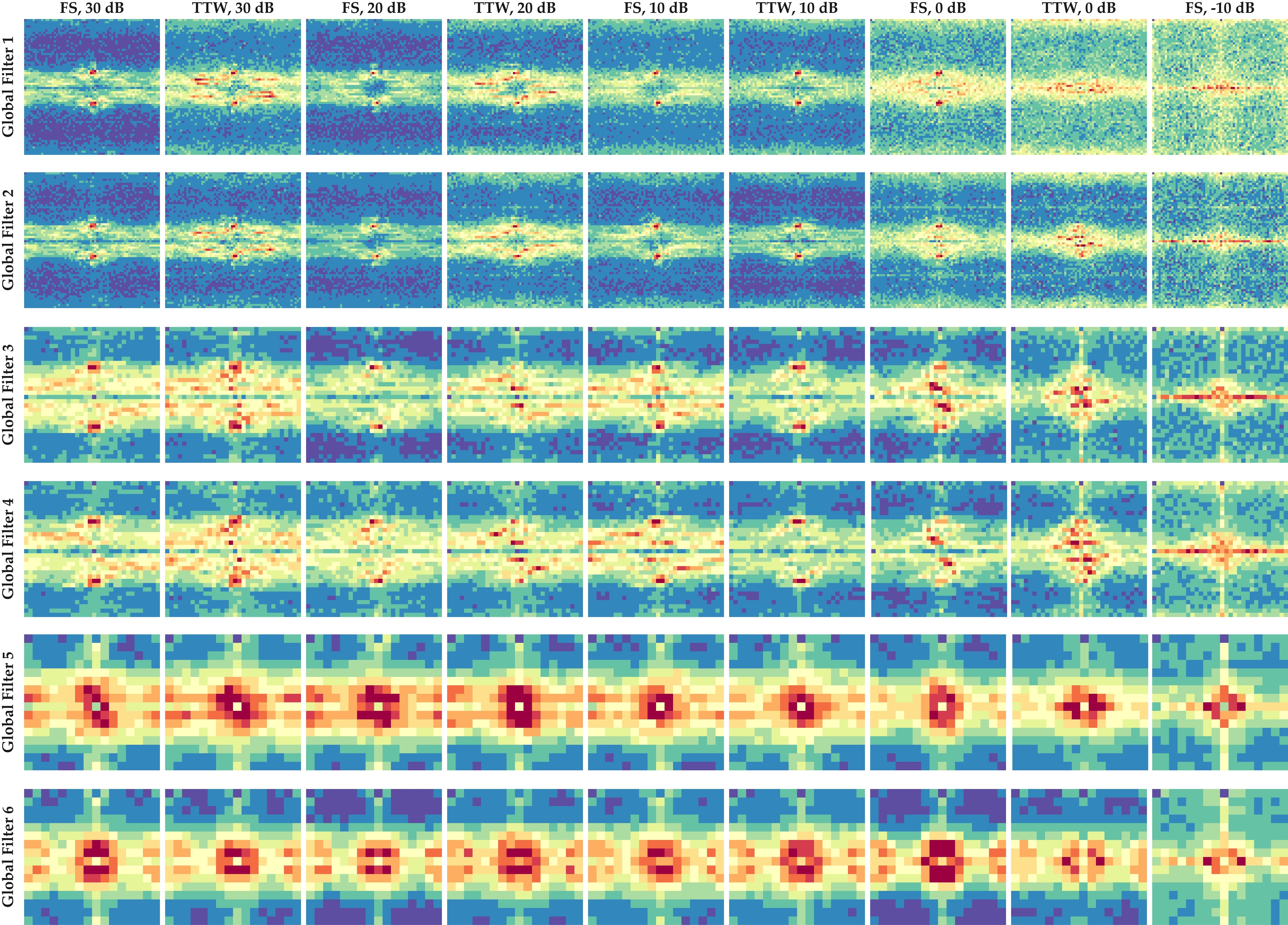}
    \caption{Weight heatmap for different layers of global filter module. The column represents the Doppler axis, and the row represents the time axis.}
    \label{Weight}
    \vspace{-0.2cm}
\end{figure*}\par
The feature separability of different activities is compared in the paper, as shown in Fig. \ref{T_SNE_FS} and \ref{T_SNE_TTW}. For STFT-based DTMs in free-space and through-the-wall scenarios under $30\mathrm{~dB}$ SNR conditions, feature dimension reduction is performed using the T-distributed stochastic neighbor embedding (T-SNE) algorithm \cite{T-SNE}. Additionally, the T-SNE feature separability before the fully connected layer of the network in the two types of scenarios is also compared. In Fig. 9, the feature separability of original DTMs for different activities is low, whereas a good feature clustering effect is achieved after neural network feature extraction. Features between different activities are clearly distinguishable. Similar conclusions can be drawn for the through-the-wall scenario in Fig. \ref{T_SNE_TTW}. This conclusion is regarded as a prerequisite for effective HAR.\par
The confusion matrices for the classification performance on the validation set are presented in Fig. \ref{Confusion_Matrics}, detailing the free-space and through-the-wall detection scenarios across various SNR levels, with entries representing the number of true positives, false positives, true negatives, and false negatives for in-place ($S1 \sim S7$) versus dynamic ($S8 \sim S12$) activities. In free-space scenario, higher SNR conditions exhibit near-perfect diagonal dominance, with minimal off-diagonal elements indicating accuracies exceeding $95\%$ and low misclassification rates, attributable to clear micro-Doppler signature that enable precise discrimination between activity types. At lower SNR, increased false positives and negatives are observed, particularly confusing subtle stationary motions with noise-induced artifacts, resulting in accuracies around $80-85\%$ due to diminished micro-Doppler resolution. Conversely, the through-the-wall matrices reveal substantial degradation, with broader off-diagonal spreads across all SNR levels, where even at $30\mathrm{~dB}$, accuracies hover below $90\%$ owing to wall-induced attenuation and noise that obscure limb signatures. At SNR of $0\mathrm{~dB}$ and $-10\mathrm{~dB}$, misclassifications intensify, with dynamic activities often erroneously categorized as stationary amid heightened noise, which yields accuracies of $65\sim 75\%$ and underscoring the challenges of propagation losses, though the overall structure preserves some separability between classes.\par

\subsection{Comparison Experiments}
The performance of the proposed network and various existing methods is compared in TABLE \ref{Comparison} with a focus on validation accuracy, model parameters, and inference time across free-space $30\mathrm{~dB}$ SNR and through-the-wall $30\mathrm{~dB}$ SNR scenarios. In the free-space scenario, the proposed method achieves a validation accuracy of $100.00\%$, which equals the performance of ViT and ConvNeXt. The proposed method requires significantly fewer parameters at $6.50\mathrm{~M}$ and a reduced inference time of $0.22~s$ compared to ViT with $85.24\mathrm{~M}$ parameters and $0.80~s$ seconds or ConvNeXt with $28.24\mathrm{~M}$ million parameters and $0.60~s$. In the through-the-wall scenario, the proposed method attains a validation accuracy of $99.44\%$, which matches ConvNeXt and surpasses other models. This method maintains its advantage in parameter efficiency and inference speed. The proposed method is demonstrated to provide an optimal balance of high accuracy, computational efficiency, and low latency. This balance makes it highly suitable for real-time applications in both scenarios.\par

\subsection{Feature Weighting Verifications}
The weight heatmaps \cite{Weighting} of various global filter layers and different SNR value under free-space and through-the-wall scenarios are comprehensively evaluated in Fig. \ref{Weight}. STFT-based DTMs are still employed for network training. Global Filter $1 \sim 6$ represent the global filter layers in the network in sequence. Global Filter $6$ shows the clearest feature followed by Global Filter $5$. Global Filter $4$, $3$, and $2$ exhibit moderate performance while Global Filter $1$ yields the worst. In free-space scenario, better classification accuracy is achieved compared to through-the-wall scenario when facing the same SNR value. This is because the network's ability to progressively remove noise during extraction is constrained by the degree of feature distortion in the image and the limitations of distinguishable units. The through-the-wall scenario is consistently more challenging for the network to learn appropriate weights than free-space conditions at corresponding SNR values. This analysis confirms the importance of both filter selection and scenarios in achieving robust HAR.\par

\subsection{Discussions}
Although the effectiveness of the proposed method is validated through various approaches such as visualization, accuracy, robustness, feature embedding, and weight heatmaps, there remain aspects worthy of improvement in the model and program design for the simulator, as well as in the architecture of the proposed neural network, including:\par
\textbf{(1) Waveform and Array Design for the Simulator:} The proposed simulator is only designed to utilize FMCW waveforms for transmission and reception, and supports simulation exclusively for single-channel stationary radar platforms. Future versions of the simulator could consider incorporating additional ultra-wideband waveforms \cite{Waveforms}, adding support for array or distributed radar systems \cite{Array}, as well as introducing functionality for moving platforms \cite{MovingPlatforms}.\par
\textbf{(2) Complex Kinematic and Echo Models:} The proposed simulator only considers $12$ types of human activities in the form of limb-node movements, and its signal model is based on simple coherent superposition. Future versions of the simulator could incorporate more complex kinematic modeling \cite{KinematicModeling}, more detailed activity configurations, and more sophisticated multipath propagation models \cite{Array}.\par
\textbf{(3) More Efficient Network Design:} The proposed network utilizes classical convolution and FFT-based global filtering operations for feature extraction and recognition. A more comprehensive network design could consider incorporating more refined signal-domain transformation strategies, attention mechanisms \cite{AttentionMechanism}, and the introduction of meta-learning to enhance the model's robustness and generalization capability \cite{MetaLearning}.\par

\section{Conclusion}
To address the difficulty for obtaining diverse and high-fidelity radar datasets for robust algorithm development, a model-based FMCW radar HAR simulator has been developed. An anthropometrically scaled $13$-scatterer kinematic model has been integrated to simulate $12$ distinct types of human activities. The FMCW radar echo model has been employed, which incorporates dynamic RCS for both free-space and through-the-wall propagation conditions, and a calibrated noise floor to ensure high signal fidelity. The simulated raw data has been processed through a complete pipeline that includes MTI, bulk Doppler compensation, and Savitzky–Golay denoising. This has been followed by the generation of high-resolution RTMs and DTMs using both the STFT and the FSST. Finally, a novel neural network method has been proposed to validate the effectiveness of the radar HAR system. Numerical experiments have demonstrated that high-fidelity and distinctive micro-Doppler signature has been successfully generated by the simulator, providing a valuable tool for the design and validation of radar HAR algorithms.\par

\section{Acknowledgement}
From the very beginning of my research, I had planned to develop simulation software. This simulator consumed nearly a full year of my spare time. It involved countless hours of grueling debugging, but thankfully, I persevered and successfully made it.\par
I would like to thank my mentors for the platform they have provided me. I would also like to thank my fellow Xiaolong Sun and junior mate Jiarong Zhao. It was their encouragement that kept me going and enabled me to complete this work.\par
My software has not undergone extensive testing by a large number of users. There may still be areas for improvement during use. I welcome your valuable feedback and would be very grateful!\par


\end{document}